\documentclass[onecolumn, draftclsnofoot]{IEEEtran}

\usepackage{setspace}
\usepackage[noadjust]{cite}
\usepackage{multirow}
\usepackage{dsfont}
\usepackage[latin1]{inputenc}
\usepackage{times}
\usepackage [spanish,english] {babel}
\usepackage [T1]{fontenc}
\usepackage [latin1]{inputenc}
\usepackage[dvips]{graphicx}
\usepackage{grffile}
\usepackage{amssymb,amsmath,amsthm, amsfonts}
\usepackage{xspace}
\usepackage{cases}
\usepackage{lineno}

\ifCLASSOPTIONcompsoc
\usepackage[tight,normalsize,sf,SF]{subfigure}
\else
\usepackage[tight,footnotesize]{subfigure}
\fi

\usepackage{algpseudocode}	
\usepackage{algorithm}
\usepackage[nolist]{acronym}

\usepackage[usenames,dvipsnames]{color}

\usepackage{pifont}

\newtheorem{lemma}{Lemma}

\newtheorem{theorem}{Theorem}

\theoremstyle{remark}
\newtheorem{remark}{Remark}

\newtheorem{prop}{Property}

\newenvironment{sequation}
{\noindent\par \vspace{-0em}\small\begin{equation}}
{\vspace{-0.8em}\end{equation}\normalsize\ignorespacesafterend}

\newenvironment{ssequation}
{\noindent\par \vspace{-0em}\footnotesize\begin{equation}}
{\vspace{-0.8em}\end{equation}\normalsize\ignorespacesafterend}

\newenvironment{seqnarray}
{\noindent \par \vspace{-2em} \small \begin{eqnarray}}
{\end{eqnarray}\normalsize\ignorespacesafterend}

\newenvironment{sseqnarray}
{\noindent \par \vspace{-0.5em} \footnotesize \begin{eqnarray}}
{\end{eqnarray}\normalsize\ignorespacesafterend}

\newcommand{\noiseVar}[1][i]{\sigma^2}

\newcommand{\WL}{water-level\xspace}
\newcommand{\WLs}{water-levels\xspace}
\newcommand{\Pool}[1][j]{\tau_{#1}}
\newcommand{\Epoch}[1][j]{\mathcal E_{#1}} 

\newcommand{\tickYes}{\textcolor{green}{\ding{52}}\xspace}
\newcommand{\tickNo}{\textcolor{red}{\ding{56}}\xspace}

\def \ie{, i.e.,\xspace}
\def \eg{, e.g.,\xspace}
\def \wrt{\ac{w.r.t.}\xspace}

\def \qos[#1]{\ac{QoS}#1}
\def \mimo[#1]{\ac{MIMO}#1}
\def \siso[#1]{\ac{SISO}#1}
\def \awgn[#1]{\ac{AWGN}#1}
\def \mmse[#1]{\ac{MMSE}#1}
\newcommand{\Precoder}[1][n]{\mat B_#1}
\newcommand{\PrecoderV}[1][n]{\mat V_{\mat B_#1}}
\newcommand{\PrecoderU}[1][n]{\mat U_{\mat B_#1}}
\newcommand{\PrecoderD}[1][n]{\mat \Sigma_{#1}}

\newcommand{\Nt}[1][n]{\ensuremath{N_t}}
\newcommand{\Nr}[1][n]{\ensuremath{N_r}}

\newcommand{\ones}[1][K]{\ensuremath{\vec 1_{#1}}\xspace}
\newcommand{\pv}[1][n]{\ensuremath{\vec p_{#1}}\xspace}
\newcommand{\MS}[1][K]{\ensuremath{\mat S_{#1}}\xspace}
\newcommand{\MMSE}[1][n]{\ensuremath{\mat E_{#1}}\xspace}

\newcommand{\Channel}[1][n]{\mat H_#1}
\newcommand{\ChannelV}[1][n]{\mat V_{\mat H_#1}}
\newcommand{\ChannelU}[1][n]{\mat U_{\mat H_#1}}
\newcommand{\ChannelD}[1][n]{\mat \Delta_{#1}}

\newcommand{\MWFlow}[1][\textit]{#1{MIMO Mercury Water-Flowing}\xspace}
\newcommand{\MWFA}{\ensuremath{\mathrm{\mathcal H_gWFA}}\xspace}
\newcommand{\MWF}{\ensuremath{\mathrm{\mathcal H_g}}WF\xspace}

\newcommand{\CMWF}{\ensuremath{C_{\MWFA}}\xspace}
\newcommand{\CCMWF}{CC_{\MWF}}

\newcommand{\DomainX}[1][]{\ensuremath{\mathcal D_{\vec x}}}

\newcommand{\Ts}[1][j]{\ensuremath{T_s}\xspace}

\def\mb{\mathbf}

\def\bs{\boldsymbol}

\def\ss#1{{\sf #1}}

\def\vec#1{\mb{#1}}
\def\vecs#1{\bs{#1}}

\def\mat#1{\mb{\uppercase{#1}}}

\def\Tr{\ss{Tr}}
\def\var{\ss{var}}

\def\diag{\ss{diag}}
\def\Diag{\ss{Diag}}
\def\sign{\ss{sign}}
\def\log{\ss{log}}
\def\rank{\ss{rank}}

\def\T{\ss{T}}
\def\d{\textrm{d}}
\def\D{\ss{D}}
\def\v{\ss{vec}}

\renewenvironment{sequation}
{\begin{equation}}
{\end{equation}\ignorespacesafterend}

\renewenvironment{seqnarray}
{\begin{eqnarray}}
{\end{eqnarray}\ignorespacesafterend}

\renewcommand{\MWFlow}[1][\textit]{#1{Mercury Water-Flowing}\xspace}

\addto\captionsenglish{\renewcommand{\figurename}{Fig.\xspace}}
\begin{document}

\title{On the precoder design  of a  wireless  energy harvesting node 
in linear vector  Gaussian channels  with  arbitrary input distribution} 

\author{{\IEEEauthorblockN{Maria Gregori and Miquel Payaró}}\\
\IEEEauthorblockA{Centre Tecnol\`ogic de Telecomunicacions de Catalunya (CTTC)\\
08860 - Castelldefels, Barcelona, Spain %. MP: He tret el punt
\\
E-mails: \{maria.gregori, miquel.payaro\}@cttc.cat \thanks{This work was partially supported by: the Catalan Government under grants 2009SGR 1046 and
2011FI\_B 00956; the Spanish Ministry of Economy and Competitiveness under project TEC2011-29006-C03-01 (GRE3N-PHY);
 and the European Commission in the framework of the FP7 Network of Excellence in Wireless COMmunications NEWCOM\# (Grant agreement no. 318306).}} 
 
% MP: Afegit%%%
\vspace{-2em}
%%%%%%%%%%%%%%%
}
%Part of this work was published previously in M. Gregori and M. Payaró, ``Optimal power allocation for a wireless multi-antenna energy harvesting node with arbitrary
%input distribution,'' in International Workshop on Energy Harvesting for Communication (ICC12 WS - EHC), Ottawa, Canada, Jun. 2012

\maketitle
\thispagestyle{empty}

\renewcommand{\WL}{water level\xspace}

\begin{abstract}
%\vspace{-0.5em}
A \ac{WEHN} operating in linear  vector Gaussian channels with arbitrarily distributed input symbols is considered in this paper.
The precoding strategy that maximizes the mutual information along $N$ independent channel accesses is studied under non-causal knowledge of the channel state and harvested energy (commonly known as offline approach).
It is shown that, at each channel use, the left singular vectors of the precoder are equal to the eigenvectors of the Gram channel matrix. 
Additionally, an expression that relates the optimal singular values of the precoder with the energy harvesting profile through the \ac{MMSE} matrix is obtained. 
%The derivation of the optimal right singular vectors of the precoder is shown to be a difficult problem and is left for future research.
Then, the specific situation in which the right singular vectors of the precoder are set to the identity matrix is considered.
In this scenario, the optimal offline power allocation, named \MWFlow, is derived and an intuitive graphical representation is presented.
Two optimal offline algorithms to compute the  \MWFlow solution  are proposed and an exhaustive study of their computational complexity is performed. Moreover, an online algorithm is designed, which only  uses causal knowledge of the harvested energy and channel state.
Finally,  the achieved  mutual information is evaluated through simulation.

\end{abstract}	

\hyphenation{op-tical net-works semi-conduc-tor theo-re-tical set-ups}

\begin{IEEEkeywords}
Energy harvesting, mutual information, arbitrary input distribution, precoder optimization, power allocation, linear vector Gaussian channels, MMSE.
\end{IEEEkeywords}

%\newpage
\section{Introduction}

Battery powered devices are becoming broadly used due to the high mobility and flexibility provided to users. As Moore predicted in 1965 \cite{moore_law}, nodes' processing capability keeps increasing as transistors shrink year after year. However, the growth of battery capacity is slower and thus energy availability is becoming the bottleneck in the computational capabilities of wireless nodes.
% as autonomy is in the order of a few days.

 Energy harvesting, which is known as the process of collecting energy from the environment by different means (e.g. solar cells, piezoelectric generators, etc.), has become a potential technology to charge batteries and, therefore, expand the lifetime of battery powered devices (e.g., handheld devices or sensor nodes), which we refer to as \acp{WEHN}. The presence of energy harvesters implies a loss of optimality of the traditional transmission policies, such as the well-known \ac{WF} strategy \cite{cover_elementsIT_1991}, because the common transmission power constraint must be replaced by a set of \acp{ECC}, which impose that energy must be harvested before it can be used by the node.

In general, the energy harvesting process is modeled as a set of energy packets arriving to the node at different time instants and with different amounts of energy.\footnote{Observe that any energy harvesting profile can be accurately modeled by making the inter arrival times sufficiently small.} There exist two well established approaches for the design of optimal transmission strategies, namely,  online and offline.
The \textit{online} approach assumes that the node only has some statistical knowledge of the dynamics of the energy harvesting process, which can be realistic in practice.
The \textit{offline} approach assumes that the node has full knowledge of the amount and arrival time of each energy packet, which is an idealistic situation that provides analytical and intuitive solutions and, therefore, it is a good first step to gain insight for the later design of the online transmission strategy.
Using this model for the energy harvesting process, references \cite{yang_optimalPacket_2010,tutuncuoglu_optimum_2010,gregori_globecom,ho_optimal_2011,Ozel_Ulukus_Yener_TX_Fading_Optimal_2011, ozel_BC_finiteBattery_2012,Ozel_BC_parallel_2012, Akif_Broadcasting_2011} 
derived the optimal resource allocation in different scenarios.
%In \cite{yang_optimalPacket_2010}, two different situations are analyzed and solved for the case of having a single-antenna \ac{WEHN} with an infinite battery capacity under an \textit{offline} approach: (i.) The node has all the data to be transmitted available from the beginning. (ii.) The data and energy packets arrive dynamically to the node. Finite battery capacity \acp{WEHN} have been studied in \cite{tutuncuoglu_optimum_2010} and \cite{gregori_globecom} solving the situations in (i.) and (ii.), respectively, for the finite battery case.
The authors of \cite{ho_optimal_2011} and \cite{Ozel_Ulukus_Yener_TX_Fading_Optimal_2011} found the power allocation strategy, named \ac{DWF}, that maximizes the total throughput of a \ac{WEHN} operating in a  point-to-point link\ie
%\small
%\vspace{-0.6em}
\begin{sequation}
\label{Eq:Directional_WF}
\sigma_n^2 = \left(W_j - \frac{1}{\lambda_n}\right)^+,\quad n \in \Pool,
\end{sequation}
%\normalsize
where $n$ is the channel use index, $\lambda_n$  is the channel gain, $\Pool$ is a set that contains the channel accesses between two consecutive energy arrivals, and $W_j$ is the water level associated to $\Pool$.\footnote{\textbf{Notation:} Matrices and vectors are denoted by upper and lower case bold letters, respectively. 
$[\vec v]_n $denotes the $n$-th component of the vector $\vec v$.
$[\mat A]_{pq}$ is the component in the $p$-th row and $q$-th column of matrix $\mat A$.
 $\mat I_n$ is the identity matrix of order $n$, \ones[n] is a column vector of $n$ ones.
$\D_{\mat X} \mat F$ denotes the Jacobian of the matrix function $\mat F$ \ac{w.r.t.} the matrix variable $\mat X$ \cite{Magnus1999}. The superscript $(\cdot)^\T$ denotes the transpose operator.
$\diag(\mat x)$ is the column vector that contains the diagonal elements of the matrix $\mat A$.
$\Diag(\vec v)$ is a diagonal matrix where the entries of the diagonal are given by the vector $\vec v$.
$\v (\mat X)$ returns a vector that stacks the columns of $\mat X$.
$\Tr (\cdot)$ denotes the trace of a matrix. $\var\{\cdot\}$ is the variance of some random variable.
The Kronecker product is denoted by the symbol $\otimes$.
Finally, $(x)^+ = \max\{0, x\}$.}

For nodes without energy harvesting capabilities, the capacity of linear vector Gaussian channels was given in \cite{telatar_MIMOcapacity_1999}, where it was shown that given a certain constraint in the transmitted power, $P_C$, capacity is achieved by diagonalizing the observed channel in $K$ independent streams. Then, a Gaussian distributed codeword is transmitted over each stream whose power is obtained from the well known \ac{WF} solution:
\begin{sequation}
\label{Eq:Waterfilling}
\sigma_k^2 = \left(W - \frac{1}{\lambda_k	}\right)^+.
\end{sequation}
The main difference between \ac{DWF} \eqref{Eq:Directional_WF} and \ac{WF} \eqref{Eq:Waterfilling} is that in the former the water level depends on the channel access under consideration.

%Let us start by summarizing the existing literature for multiple-antenna non-harvesting nodes and afterwards some pioneering works in optimal transmission strategies for single-antenna \acp{WEHN}.

Both \eqref{Eq:Directional_WF} and \eqref{Eq:Waterfilling} are optimal when the distribution of the input is Gaussian. However, in practical scenarios, finite constellations are used instead of the ideal Gaussian signaling\eg $Q$-PAM and $Q$-QAM, where $Q$ denotes the alphabet cardinality. In the low \ac{SNR} regime, the capacity achieved with finite constellations is very close to the one achieved by Gaussian signaling. However, the mutual information asymptotically saturates when the \ac{SNR} increases as not more than $\log_2 Q$ bit per channel use can be sent (see Fig. 1 in \cite{forney_modulation_1998}). 
This must be taken into account in the design of the optimal power allocation when the input symbols are constrained to belong to a finite alphabet. In opposition to the Gaussian case, where the better the channel gain, the higher the allocated power, when arbitrary constellations are used, there exists a tradeoff between the alphabet cardinality and the channel gain. 
In \cite{Lozano_mercuryWF_2006}, the optimal power allocation was found for a node without energy harvesting capabilities and with arbitrary distributed input symbols. To do so, the authors of \cite{Lozano_mercuryWF_2006} used the relation between the mutual information and the \ac{MMSE}, which was revealed in \cite{Guo_I_mmse_2005} and further generalized in \cite{palomar_gradientMI_2006}, as summarized in the following lines.

In \cite{Guo_I_mmse_2005}, Guo et al. revealed that the derivative of the mutual information with respect to (w.r.t.) the \ac{SNR} for a real-valued scalar Gaussian  channel is proportional to the \ac{MMSE}\ie
\begin{sequation}
\label{Eq:I_mmse}
\frac{\d}{\d snr} I(x; \sqrt{snr}  x + n) = \frac{1}{2}mmse(snr),
\end{sequation}
where $x$ is the channel input, $n$ is the observed noise and $mmse(snr)= \mathbb{E}\:\{(x - {\hat{x}})^2 \}$, where $\hat{x} = \mathbb{E}\:\{ x | \sqrt{snr}  x + n \}$ is the conditional mean estimator.
The mutual information in linear vector Gaussian channels was further characterized in \cite{palomar_gradientMI_2006}, where its partial derivatives  \ac{w.r.t.} arbitrary system parameters were determined\eg the gradient  \ac{w.r.t.} the channel matrix, $\mat H$, was found to be 
\begin{sequation}
\label{Eq:I_mmse_Palomar}
\nabla_{\mat H} I(\vec x; \mat H \vec x + \vec n) = \mat H \mat E,
\end{sequation}
where $\vec x$ is the vector input, $\vec n$ is the noise, $\mat E = \mathbb{E}\:\{(\vec x - \vec{\hat{x}})(\vec x - \vec{\hat{x}})^\T \}$ is the \ac{MMSE} matrix, and  $\vec{\hat{x}} = \mathbb{E}\:\{ \vec x | \mat H \vec x + \vec n \}$.

Thanks to the relationship in \eqref{Eq:I_mmse}, the power allocation that maximizes the mutual information over a set of parallel channels (each of them denoted by a different index $k$) with finite alphabet inputs was derived in \cite{Lozano_mercuryWF_2006} and named Mercury/Waterfilling (\MWF)\ie
\begin{sequation}
\label{Eq:Lozano_mercury}
\sigma_k^2 = \left(W - \frac{1}{\lambda_k}G_k\left(\frac{1}{W\lambda_k}\right)\right)^+,
\end{sequation}
where
 $G_k(\psi)$ is the mercury factor that depends on the input distribution and is defined as
 \begin{sequation}
\label{Eq:G}
G_k(\psi) = \left\{ \,
\begin{IEEEeqnarraybox}[][c]{l?s}
\textstyle
\IEEEstrut
 \frac{1}{\psi} - mmse_k^{-1}\left( \psi  \right) & if $\psi \in [0,1]$, \\
1 & if $\psi \geq 1.$ 
\IEEEstrut
\end{IEEEeqnarraybox}
\right. 
\end{sequation}
This result showed that the optimal power allocation not only depends on the channel gain as in the Gaussian signaling case, but also on the shape and size of the constellation. 
%The \MWF solution assumes a diagonal channel matrix and no precoding at the transmitter.

% The effect of the linear precoding is studied in \cite{Perez_MIMO_Precoding_arbitrary_2010} showing that in general a non-diagonal precoding matrix increases the mutual information for non Gaussian input distributions.

The goal of this work is to design the transmitter that maximizes the mutual information along $N$ channel uses by 
jointly considering the nature of the energy harvesting process at the transmitter and arbitrary distributions of the input symbols, which, to the best of our knowledge, has not been yet considered in the literature. 
Hence, the main contributions of this paper are:
(i.) Proving that, at the $n$-th channel use, the left singular vectors of the $n$-th precoder matrix are equal to the eigenvectors of the $n$-th channel  Gram matrix.
(ii.) Deriving an expression that relates the singular values of the $n$-th precoder matrix with the energy harvesting profile through the \ac{MMSE} matrix.
(iii.) Showing that the derivation of the optimal right singular vectors is a difficult problem and proposing a possible research direction towards the design of a numerical algorithm that computes the optimal right singular vectors. 
The design of this numerical algorithm is out of the scope of the current paper because our focus is to gain insight from the closed form power allocation that is obtained after setting the right singular vectors matrix to be the identity matrix
and, in this scenario, the contributions are: 
(iv.) Deriving the optimal  offline power allocation, named the \MWFlow solution, and providing an intuitive graphical interpretation, which follows from demonstrating that the mercury level is monotonically increasing with the water level.
(v.) Proposing two different algorithms to compute the \MWFlow solution, proving their optimality, and carrying out an exhaustive study of their computational complexity.
(vi.) Implementing an online algorithm, which does not require future knowledge of neither the channel state nor the energy arrivals, that computes a power allocation that performs close to the offline optimal \MWFlow solution.

The remainder of the paper is structured as follows. Section \ref{Sec:System_model} presents the system model. In Section \ref{Sec:Problem}, the aforementioned problem is formally formulated and solved. The graphical interpretation of the \MWFlow solution is given in Section \ref{Sec:MWFlow}. The offline and online algorithms are introduced in Sections  \ref{Sec:Algorithms} and 
\ref{S:online}, respectively.
In Section \ref{Sec:Results}, the performance of our solution is compared with different suboptimal strategies and the computational 
complexity of the algorithms is experimentally evaluated. Finally, the paper is concluded in Section \ref{Sec:Conclusions}.

%Section \ref{Sec:System_model} presents the system model. In Section \ref{Sec:Problem}, the aforementioned problem is formally formulated and solved. An interpretation of the solution, named \textit{\ac{MIMO} Mercury Waterflowing}, is given in Section \ref{Sec:Interpretation}. The developed algorithm is presented in Section \ref{Sec:Algorithm}. In Section \ref{Sec:Results}, the performance of our solution is compared with different suboptimal strategies. Finally, the paper is concluded in Section \ref{Sec:Conclusions}.

\section{System model}
\label{Sec:System_model}
We consider a point-to-point communication through a discrete-time linear vector Gaussian channel where the transmitter is equipped with energy harvesters.
%\ie the transmitter is a \ac{WEHN}.
% Similarly, $n_R$ denotes the number of antennas at the receiver.
A total of $N$ channel uses are considered where at each channel use the symbol $\vec s_n \in \Re^L$ is transmitted.\footnote{
The real field has been considered for the sake of simplicity. 
The extension to the complex case is feasible but requires 
the definition of the complex derivative, the generalization of the chain rule, and cumbersome mathematical derivations, which is out of the scope of this work. 
Nevertheless, the extension to the complex case can be done similarly as \cite{xiao_globallyOptimal_2011} generalized the results obtained in \cite{Payaro_Hessian_2009}.}

\begin{figure}[t]
\centering
\includegraphics[width=0.8\columnwidth]{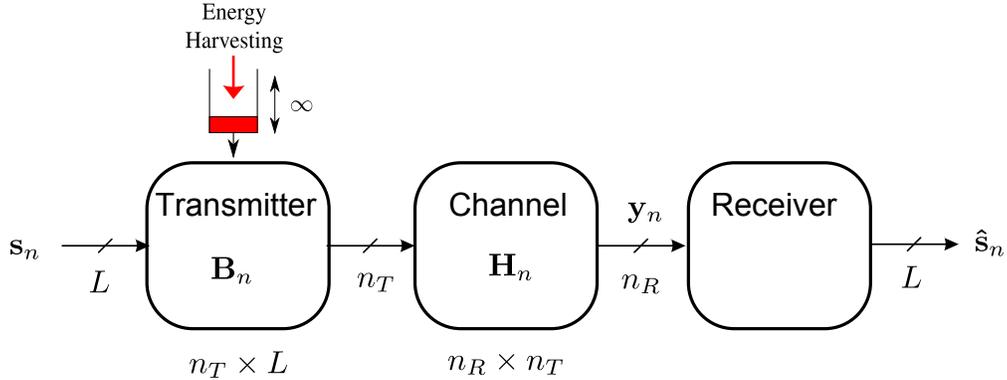}
\caption{The discrete-time linear vector Gaussian channel at the $n$-th channel use.}
\label{fig:block_general}
\vspace{-1em}
\end{figure}

We consider that the symbols $\{\vec s_n \}_{n=1}^N$ have independent components with unit power\ie $\mat r_s = \mathbb{E}\{\vec s_n \vec s_n^\T\}= \mat I_L$ and that they are \ac{i.i.d.} along channel uses according to $P_S(\vec s_n)$. As shown in \figurename \ref{fig:block_general}, the symbol $\vec s_n$ is linearly processed at the transmitter by the precoder matrix  $\Precoder \in \Re^{\Nt \times L}$. We consider a slow-fading channel where the coherence time of the channel $T_C$ is much larger than the symbol duration $\Ts$\ie $\Ts \ll T_C$. 
Thus, a constant channel matrix $\mat H_n  \in \Re^{\Nr \times \Nt}$ is considered at the $n$-th channel use.
Let $K$ denote the rank of the channel matrix\ie $K= \rank(\Channel)=\min\{\Nt,\Nr\}$, then we have that $L \leq K$.\footnote{
We have considered that $\Channel$ is not rank deficient, $\forall n$, which is a realistic assumption due to random nature of the channel.}
%\textcolor{red}{, where the entry $[\mat H_n]_{pq}$ contains the observed channel gain between the $q$-th antenna of the transmitter and the $p$-th receiver's antenna}. 
%$\mat R_{H_n} =\mat H_n^\T \mat H_n$ is the channel Gram matrix.
Thus, the received signal at the $n$-th channel use is
\begin{equation}
\label{Eq:Sys_model}
\vec y_n = \Channel \Precoder \vec s_n  + \vec{w}_n,
\end{equation}
where $\vec{w}_n$ represents the zero-mean Gaussian noise with identity covariance matrix $\mat R_{w_n} = \mat I_{\Nr}$.\footnote{
Note that if the noise is colored and its covariance matrix $\mat R_{w_n}$ is known, we can consider the whitened received signal $\mat R_{w_n}^{-1/2} \vec y_n$.} Let $\MMSE$ denote the $n$-th channel use \ac{MMSE} matrix, which is defined as $\MMSE = \mathbb{E}\:\{(\vec s_n - \vec{\hat{s}}_n)(\vec s_n - \vec{\hat{s}}_n)^\T \}$ and  $\vec{\hat{s}}_n = \mathbb{E}\:\{ \vec s_n | \vec y_n \}$ is the conditional mean estimator.

Let us express the channel matrix as $\Channel = \ChannelV \ChannelD \ChannelU^\T$, where $\ChannelD \in \Re^{L \times L}$ is a diagonal matrix that contains the $L$ largest eigenvalues of  $\Channel$ and 
$\ChannelV\in \Re^{\Nr \times L}$ and $\ChannelU \in \Re^{\Nt \times L}$ are semi-unitary matrices that contain the row and column associated eigenvectors, respectively.
The precoder matrix $\Precoder$ can be expressed as $\Precoder = \PrecoderU \PrecoderD \PrecoderV^\T$, where $\PrecoderU\in \Re^{\Nt \times L}$, $\PrecoderD\in \Re^{L \times L}$ is a diagonal matrix whose entries are given by the vector  $\vecs \sigma_n = [\sigma_{1,n}, \dots ,\sigma_{L,n}]^\T$ and $\PrecoderV\in \Re^{L \times L}$ is a unitary matrix.
Full \ac{CSI} is assumed at the transmitter. 

\begin{figure}
\centering
\includegraphics[width=0.6\columnwidth]{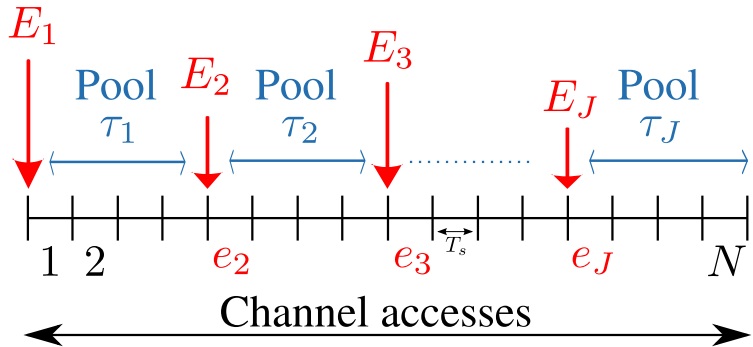}
\caption{Temporal representation of energy arrivals.}
%\vspace{-1em}
\label{fig:temporal}
%\vspace{-1em}
\end{figure}

The energy harvesting process at the transmitter is characterized by a packetized model\ie the node is able to collect a packet of energy containing $E_j$ Joules at the beginning of the $e_j$ channel access. Let $J$ be the total number  of packets harvested during the $N$ channel uses. The initial battery of the node is modeled as the first harvested packet $E_1$ at $e_1=1$. 
We assume that the mean time between energy arrivals, $T_e$, is considerably larger than the symbol duration time\ie $T_e \gg \Ts$ and thus we can consider that packet arrival times are aligned at the beginning of a channel use.\footnote{
In our model, the transmitter can only change its transmission strategy in a channel access basis. Accordingly, if an energy packet arrives in the middle of a channel access, we can assume that the packet becomes available for the transmitter at the beginning of the following channel access.}
First, in Sections \ref{Sec:Problem}-\ref{Sec:Algorithms}, we consider the offline approach as it provides analytical and intuitive expressions. Afterwards, in Section \ref{S:online}, we develop an online transmission strategy where the transmitter only has causal knowledge of the energy harvesting process\ie about the past and present energy arrivals.
We use the term \textit{pool}, $\tau_j, j = 1 \dots J$, to denote the set of channel accesses between two consecutive energy arrivals.
As in \cite{yang_optimalPacket_2010} and \cite{ozel_BC_finiteBattery_2012}, we assume an infinite capacity battery  since, in general, the battery size is
large enough so that the difference between the accumulated harvested energy and the accumulated expended energy is always smaller than the battery capacity.
A temporal representation  is given in \figurename \ref{fig:temporal}.

%In next section, the problem of designing the set of linear precoder matrices, $\{\Precoder\}_{n=1}^N$, that maximizes the mutual information is formally formulated and solved.

%\vspace{-1em}
\section{Throughput maximization problem}
\label{Sec:Problem}
% we derive the set of precoding matrices $\{\Precoder\}_{n=1}^N$ that maximize the sum of the mutual information between input and output along the $N$ time slots\ie $I=\sum_{n=1}^{n=N} I(\vec s_n, \vec y_n)$, where $I(\vec s_n, \vec y_n)$ represents the mutual information of the $n$-th channel use. The optimal design must satisfy a set of \textit{energy causality} constraints\ie the energy cannot be used before it has been harvested.
In this section, we study the set of linear precoding matrices $\{\Precoder\}_{n=1}^N$ that maximizes the input-output mutual information along $N$ independent channel accesses, $\sum_{j=1}^{J} \sum_{n \in \Pool} I(\vec s_n; \vec y_n)$, where $I(\vec s_n; \vec y_n)$ is the $n$-th channel use mutual information.
The design of $\{\Precoder \}_{n=1}^N$ is constrained to satisfy instantaneous \acp{ECC}, which impose that energy cannot be used before it has been harvested, $\Ts \sum_{j=1}^\ell \sum_{n\in \Pool} || \Precoder \vec s_n ||^2 \leq \sum_{j=1}^\ell E_j $, $\ell =1 \dots J$.
However, since in each pool there are several channel accesses with the same channel gains (because $T_c \gg \Ts$ and $T_e\gg \Ts$), 
instead of imposing the instantaneous \acp{ECC}, 
we can consider the mean \acp{ECC} that become $\Ts  \sum_{j=1}^\ell \sum_{n\in \Pool} \Tr (\Precoder \Precoder^\T) \leq \sum_{j=1}^\ell E_j $, $\ell =1 \dots J$,
which do not require prior knowledge of the transmitted symbols at each channel use as only the expectation of the symbols is needed.\footnote{
In general, the energy harvesting and the channel state are two independent random processes, thus, there may be situations in which 
only a few a few channel accesses separate an energy arrival from a change of the channel realization, however, note that these situations are unlikely since $T_c \gg \Ts$ and $T_e \gg \Ts$. In these improbable situations, 
the temporal averaging is not sufficient to ensure that  the fulfillment of the mean \acp{ECC} implies a fulfillment of the instantaneous \acp{ECC}, however, the averaging through the different channel dimensions brings closer the mean and instantaneous \acp{ECC}.
Thus, the mean \acp{ECC} can be used instead of the instantaneous  \acp{ECC} since the cases in which they differ are indeed very unlikely.}

%\textcolor{blue}{The design of $\{\Precoder \}_{n=1}^N$ is constrained to satisfy instantaneous \acp{ECC}, which impose that energy cannot be used before it has been harvested, $\sum_{j=1}^\ell \sum_{n\in \Pool} || \Precoder \s ||^2 \leq \sum_{j=1}^\ell E_j $, $\ell =1 \dots J$.
%In each pool, there are several channel accesses with the same channel gains since $T_c \gg \Ts$ and $T_e\gg \Ts$, thus, instead of considering the instantaneous \acp{ECC}, we can consider the mean \acp{ECC} that are $ \Ts \sum_{j=1}^\ell \sum_{n\in \Pool} \Tr (\Precoder \Precoder^\T) \leq \sum_{j=1}^\ell E_j $, $\ell =1 \dots J$.}
Therefore, the mutual information maximization is mathematically expressed as
%\vspace{-1em}
\begin{subequations}
%\small
\label{Eq:P_sloted}
\begin{seqnarray}
\label{Eq:P_sloted_1}
\hspace{-1.5em} \max_{\{\Precoder\}_{n=1}^N} && \hspace{-1em}  \sum_{j = 1}^J \sum_{n \in \Pool} I(\vec s_n; \vec y_n) \\
%\end{seqnarray}
%\normalsize
%subject to
%\small
%\begin{seqnarray}
s.t. && \hspace{-1em} \Ts \sum_{j=1}^\ell \sum_{n\in \Pool} \Tr (\Precoder \Precoder^\T) \leq \sum_{j=1}^\ell E_j, \:  \: \ell =1 \dots J			.
\label{Eq:P_sloted_3}
\end{seqnarray}
\end{subequations}
%\normalsize
%where $H(\cdot)$ is the Heaviside function with $H(0)=0$.

%Note that if $J=1$ and $e_1 = 1$ in \eqref{Eq:P_sloted}, then the obtained problem deals with the design of the precoding strategy that maximizes the mutual information for a node without energy harvesting capabilities.
%If the node does not harvest energy 
%bserve that if the node does not harvest energy over time\ie $J=1$ and $e_1 = 1$, the Problem in \eqref{Eq:P_sloted} studies .ma
Before addressing the problem in \eqref{Eq:P_sloted}, let us summarize the state of the art on the precoding strategy that maximizes the mutual information for non-harvesting nodes, which was studied in \cite{Payaro_Palomar_Optimal_2009, Payaro_Hessian_2009,  Lamarca_2009_Precoding, xiao_globallyOptimal_2011} and references therein.\footnote{
When there is no energy harvesting in the transmitter, the mutual information maximization problem is the one obtained after setting $J=1$ and $N =1$ in \eqref{Eq:P_sloted}. 
Thus, the mutual information is maximized for a single channel use under a power constraint.}
In \cite{Payaro_Palomar_Optimal_2009}, it was shown that, in general,
the mutual information,  $I(\vec s_n; \vec y_n)$,  is not a concave function of the precoder and that depends on the precoder only through the matrix $\mat Z_n = \Precoder^\T \Channel^\T \Channel \Precoder$.
The authors of \cite{Payaro_Palomar_Optimal_2009} also showed that the left singular vectors of the precoder can be chosen to be equal to the eigenvectors of the channel Gram matrix\ie $\PrecoderU = \ChannelU$. From this, $\mat Z_n = \PrecoderV \PrecoderD^2 \ChannelD^2  \PrecoderV^\T$ and the  mutual information depends on the precoder only through the right eigenvectors and the associated singular values.
In \cite{Payaro_Hessian_2009}, it was shown that $I(\vec s_n; \vec y_n)$  is a concave function of the squared singular values of the precoder, $\diag(\PrecoderD^2)$, when a  diagonal channel matrix is considered. Finally, the authors of \cite{Payaro_Palomar_Optimal_2009} stated that the complexity in the design of the globally optimal precoder lies in the right singular vectors of the precoder, $\PrecoderV$.
Then, in \cite{Lamarca_2009_Precoding}, it was shown that  $I(\vec s_n; \vec y_n)$ is a concave function of the matrix $\mat Z_n$ and a gradient algorithm over $\mat Z_n$ was derived to find a locally optimal precoder.
References  \cite{Payaro_Palomar_Optimal_2009, Payaro_Hessian_2009,  Lamarca_2009_Precoding} considered  a real channel model. 
The extension to the complex case was done in \cite{xiao_globallyOptimal_2011}, where the authors pointed out that by allowing the precoder and the channel matrix to be in the complex field the mutual information can be further improved. Then, they proposed an iterative algorithm that determines the globally optimal precoder that imposes that the power constraint must be met with equality.

When energy harvesting is considered,
instead of having a single power constraint, we have a set of $J$ \acp{ECC} as in \eqref{Eq:P_sloted_3} and it is not straightforward to determine which of the constraints must be met with equality. 
This fact implies that the algorithm introduced in \cite{xiao_globallyOptimal_2011} is no longer optimal when energy harvesting is considered.
Altogether, \eqref{Eq:P_sloted} is not a convex optimization problem since the mutual information is not a concave function of the precoder and, hence, its solution
is not straightforward. 
In the following lemma, we generalize Proposition 1 in \cite{Payaro_Palomar_Optimal_2009}  for the case of considering energy harvesting in the transmitter.

\begin{lemma}
\label{L:Diag}
The left singular vectors of the $n$-th precoder matrix, $\PrecoderU$, are equal to the eigenvectors of the channel Gram matrix $\ChannelU$,  $\forall n$.
\end{lemma}
\begin{IEEEproof}
See Appendix \ref{A:Proof_Diagonalize}.
\end{IEEEproof}

%\textcolor{red}{\sout{The problem of finding $\PrecoderV^{\T \star}$ has been studied for non-energy harvesting nodes in \cite{Payaro_Palomar_Optimal_2009}, where the authors showed that, for non-Gaussian inputs, the computation of $\PrecoderV^{\T \star}$ is an extremely hard problem. 
%In \cite{xiao_globallyOptimal_2011}, an iterative numerical algorithm was recently proposed to find the globally optimal precoder matrix.  A very interesting research direction is to expand the iterative algorithm developed in \cite{xiao_globallyOptimal_2011} to include energy harvesting at the transmitter. However, this is left as an open problem as we preferred to consider a fixed and known $\PrecoderV^{\T}$ because this allows us to derive an analytical and intuitive expression for the optimal singular values matrix of the precoder, $\PrecoderD^{\star}$.  
%} }

Thanks to Lemma \ref{L:Diag}, the optimal precoding matrix is $\Precoder^\star = \ChannelU \PrecoderD^\star \PrecoderV^{\T \star}, \forall n$,
and the dependence of $I(\vec s_n; \vec y_n)$ on the precoder is only through $ \PrecoderD$ and $\PrecoderV$.
In the following lines, we maximize the mutual information \acl{w.r.t.}  $\PrecoderD$ for a given $\PrecoderV$.
By applying Lemma \ref{L:Diag} in \eqref{Eq:Sys_model}, the next equivalent signal model is obtained
\begin{sequation}
\label{eq:Eq_signal_model}
\vec y_n=  \tilde{\mat H}_n   \vec s_n  +  \vec{w}_n,
\end{sequation}
where $\tilde{\mat H}_n  = \ChannelV \ChannelD \PrecoderD \PrecoderV^\T$ and $\PrecoderV^\T$ is deterministic and known.
To fully exploit the diversity of the channel, we assign the dimension of the input vector to be equal to the number of channel eigenmodes\ie$L =K$.
 It is easy to verify that 
the maximization of the mutual information \ac{w.r.t.} $\PrecoderD$ is not a convex optimization problem. However, if instead we maximize the mutual information w.r.t. the squared singular values of the precoder $\pv= [\sigma_{1,n}^2, \dots ,\sigma_{K,n}^2]^\T$, the obtained problem is convex, as shown in the following lines. Thus, the problem reduces to 
\begin{subequations}
\label{Eq:Problem_Equivalent}
\begin{seqnarray}
\hspace{-2.5em}\max_{\{\pv\}_{n=1}^N} &&  \sum_{j = 1}^J \sum_{n \in \Pool}  I(\vec s_n ;   \tilde{\mat H}_n   \vec s_n  +  \vec{w}_n)\\
s.t. &&  \Ts\sum_{j=1}^\ell \sum_{n\in \Pool}   \ones^\T \pv \leq \sum_{j=1}^\ell E_j , \quad \ell =1 \dots J. \quad			\label{Eq:Problem_Equivalent_ECC}
%\label{Eq:P_sloted_3}
\end{seqnarray}
\end{subequations}
Observe that, at the $n$-th channel access, the input-output mutual information $I(\vec s_n ;   \tilde{\mat H}_n   \vec s_n  +  \vec{w}_n)$ is concave w.r.t. $\pv$, which was proved in \cite{Payaro_Hessian_2009}. Therefore, the objective function is concave as the sum of concave functions is concave \cite{boyd_convex_2004}. Finally, as the constraints are affine in $\pv$, \eqref{Eq:Problem_Equivalent} is a convex optimization problem and the KKT are sufficient and necessary optimality conditions. In particular, the optimal solution must satisfy $\D_{\pv} \mathcal L = \mat 0$ (the reader who is not familiar with this notation, which is presented in \cite{Magnus1999}, is referred to \cite[Appendix B]{Payaro_Hessian_2009} for a concise summary), where $\mathcal L$ is the Lagrangian that is
$\mathcal L =  \sum_{j = 1}^J \sum_{n \in \Pool}   I(\vec s_n ;   \tilde{\mat H}_n   \vec s_n  +  \vec{w}_n)
 - \sum_{\ell=1}^{J} \rho_\ell \left( \Ts \sum_{j = 1}^\ell \sum_{n \in \Pool}   \ones^\T \pv - \sum_{j=1}^\ell E_j\right)$, where $\{\rho_\ell \}_{\ell = 1}^{J}$ are the Lagrange multipliers associated with the inequality  constraints.
We want to remark that in all the expressions derived in the remainder of the paper, $n$ refers to some channel access contained in $\Pool$, which follows from the formulation of $\mathcal L$.  
In order to obtain $\D_{\pv} \mathcal L$, we first need to determine the Jacobian matrix of the mutual information \wrt $\pv$, which is done in the following lemma:
%\vspace{-1em}
\begin{lemma}
\label{L:JacobianI}
The Jacobian matrix of the mutual information \wrt $\pv$ is $\D_{\pv}  I(\vec s_n ;   \tilde{\mat H}_n   \vec s_n  +   \vec{w}_n) = \frac{1}{2} \diag^\T \left( \ChannelD^2 \PrecoderV^\T \MMSE   \PrecoderV  \right)$.
%\vspace{-0.5em}
\end{lemma}
\begin{IEEEproof}
%\vspace{-0.2em}
See Appendix \ref{A:JacobianI}.
\end{IEEEproof}

With this result, we can proceed to solve the KKT condition $\D_{\pv} \mathcal L = \mat 0$:
%
%\vspace{-0.1em}
%\small
\begin{seqnarray}
\D_{\pv} \mathcal L = \frac{1}{2} \diag^\T \left( \ChannelD^2 \PrecoderV^\T \MMSE   \PrecoderV  \right) - \Ts\sum_{\ell= j}^{J} \rho_{\ell} \ones^\T  = \mat 0 \Rightarrow \nonumber \\
\Rightarrow \left[\ChannelD^2 \PrecoderV^\T \MMSE   \PrecoderV \right]_{kk} =  \frac{1}{W_j},  \quad k= 1 \dots K, n\in \Pool, \label{Eq:FirstKKT}
\end{seqnarray}
where $W_j$ is the $j$-th pool water level\ie
\begin{sequation}
\label{WQ:WL}
W_j =\frac{1}{2\Ts  \sum_{\ell= j}^{J} \rho_{\ell} }.
\end{sequation}
From \eqref{Eq:FirstKKT},  at each channel use, we obtain a set of $K$ conditions that relate the power allocation in each stream (through the \ac{MMSE} matrix) with the energy harvesting profile (through the pool's water level).
Some properties of the water level $W_j$ can be derived from the KKT optimality conditions:
\begin{seqnarray}
\rho_\ell \geq 0, &&\quad \forall \ell\label{Eq:Slackness_1},\\
\rho_\ell\big( \Ts\sum_{j=1}^\ell \sum_{n\in \Pool}  \ones^\T \pv  - \sum_{j=1}^\ell E_j \big) = 0, && \quad \forall \ell.\label{Eq:Slackness_2}
\end{seqnarray}
%\normalsize
%\vspace{-1em}
Plugging \eqref{Eq:Slackness_1} in  \eqref{WQ:WL}, it is straightforward to obtain the following property:
\begin{prop}
\label{prop:1}
The water level is non-decreasing in time.\footnote{
This property is only valid under an infinity battery capacity assumption. When a finite battery is considered the water level may increase or decrease \cite{Ozel_Ulukus_Yener_TX_Fading_Optimal_2011}.
}
\end{prop}
From \eqref{Eq:Slackness_2}, we can get more insights in the solution. There are two possibilities to fulfill \eqref{Eq:Slackness_2}:
\begin{itemize}
    \item Empty Battery: This situation occurs when, at the end of the $\ell$-th pool, the node has consumed all the energy, i.e., $ \Ts\sum_{j=1}^\ell \sum_{n\in \Pool}  \ones^\T \pv  - \sum_{j=1}^\ell E_j  = 0$. 
    \item Energy Flow: This situation occurs when, at the end of the $\ell$-th pool, the node has some remaining energy in the battery, which will be used in the following pools. When this happens $\rho_\ell =0$ and, hence, $W_{\ell+1} = W_{\ell}$.
\end{itemize}

\begin{prop}
\label{prop:emtyBatery}
Changes on the water level are only produced when at the end of the previous pool the node has consumed all the available energy. 
\end{prop}

Note that the \acp{ECC} take into account
the energy spent by the node over all the dimensions. Thus, these two properties also hold in a
scalar channel model as proved in \cite[Theorem 3]{ho_optimal_2011}.

Since the problem in \eqref{Eq:Problem_Equivalent} is convex, by using \eqref{Eq:FirstKKT} and Properties \ref{prop:1} and \ref{prop:emtyBatery}, we can construct efficient numerical algorithms to compute the optimal power allocation, $\{\pv^\star\}_{n=1}^N$, for a given $\PrecoderV$. 
The maximization of the mutual information \ac{w.r.t.} $\PrecoderV$ is indeed much more complicated as pointed out in \cite{Payaro_Palomar_Optimal_2009} for the non-harvesting scenario.
In this context, in this work, we focus on the particular case in which $\PrecoderV = \mat I_K$ because, in spite of not being necessarily the globally optimal precoder, it leads to an analytical closed form power allocation that allows an intuitive graphical representation of the solution,  as it is explained in the next section.
Observe that for any other choice  $\PrecoderV \neq \mat I_K$, we must resort to numerical methods to compute the optimal power allocation.

The design and development of a numerical algorithm that computed the globally optimal precoder at each channel access
would be an interesting research problem in its own and is left for future research. We believe that a possible starting point would be to analyze how to expand the algorithms presented in \cite{Lamarca_2009_Precoding} and \cite{xiao_globallyOptimal_2011}, which exploit the concavity of the mutual information \ac{w.r.t.} the matrix $\mat Z_n$ and the fact that the power constraint must be met with equality, to the energy harvesting scenario. 
Note that if we knew the optimal total power allocation in each channel access, we could run  $N$ times the algorithm proposed in \cite{Lamarca_2009_Precoding} to obtain the globally optimal precoder in each channel access, however, 
this approach has two major drawbacks. First, the optimal total power allocation in each channel access is not known a priori and its computation is not straightforward since the total power consumptions of the different channel accesses belonging to the same pool must simultaneously satisfy \eqref{Eq:FirstKKT}. 
The second drawback is the required computational burden since any iterative approach requires a new estimation of the \ac{MMSE} matrix, $\MMSE$, at every iteration since it depends on $\PrecoderV$ and $\PrecoderD$.
These two reasons makes challenging the applicability of the proposed approach and, hence, different alternatives to find the globally optimal precoder may be required. Altogether, we believe that the development of a numerical algorithm that computes the globally optimal precoder for a \ac{WEHN}  is the object of a new paper in its own and is left for future research.

%\textcolor{red}{As the problem in \eqref{Eq:Problem_Equivalent} is convex, there are well known and efficient algorithms to find the optimal power allocation matrix $(\PrecoderD^{2})^\star$ for a given $\PrecoderV^{\T \star}$, e.g., the interior point method \cite{boyd_convex_2004}.With this result, we extend the \ac{DWF} solution found in \cite{Ozel_Ulukus_Yener_TX_Fading_Optimal_2011} to the linear vector Gaussian channel with arbitrary input distributions.}

%By using \eqref{P:Power_Allocation_matrix} and Corollary \ref{C:WL}, it is easy to develop algorithms that efficiently compute the optimal power allocation matrix $ \Sigma_{{n}}^{2\star}$ for a given $\PrecoderV^{\T \star}$.
%\vspace{-0.5em}
\section{The \MWFlow[] solution}
\label{Sec:MWFlow}
%This choice of  $\PrecoderV^{\T} $ has been made because it allows a nice and intuitive representation of the solution, named \textit{Mercury Water-Flowing}, which is presented in the following lines.

In the remainder of the paper,  we consider a communication system in which the precoder is constrained to satisfy  $\PrecoderV = \mat I_K$ or, equivalently, a communication system such that both the precoder and channel matrices are diagonal. 
In spite of the fact that total achievable mutual information is reduced by forcing $\PrecoderV = \mat I_K$, we consider that it is interesting to study this scenario for the following three reasons: 
(i.) The system $\vec y_{n}' = \ChannelD \PrecoderD \vec s_n +  \vec{w}'_n$, with $\vec{w}'_n$ being the observed noise at the receiver, is commonly encountered in practical systems where, for simplicity at the decoder, independent symbols are transmitted in each dimension (e.g.,  in multi-tone transmissions like \ac{OFDM}), and it has been broadly considered in the literature, indeed, the \MWF solution was derived for such an input-output system model in \cite{Lozano_mercuryWF_2006}.
(ii.) The optimal power allocation, which is named \textit{Mercury Water-Flowing}, 
accepts a closed form expression and an intuitive graphical representation.
(iii.) We believe that the intuition gained thanks to the  \textit{Mercury Water-Flowing} graphical interpretation may help for the design of the algorithm that computes the globally optimal precoder of the problem in \eqref{Eq:P_sloted}.

%\textcolor{red}{In this context, 
%the input-output model $\vec y_{n}' = \ChannelD \PrecoderD \vec s_n$  can be obtained from the general model in \eqref{eq:Eq_signal_model} by setting $\PrecoderV^{\T} =  \mat I_K $ and $\vec y_{n}'=\ChannelV^\T \vec y_n$. Thus,  a set of $K$ independent parallel streams are observed at each channel use, where  the received signal in the $k$-th stream is 
%$ y_{k,n}' = \sqrt{\lambda_{k,n}\sigma_{k,n}^2} s_{k,n} + w_{k,n}',$
%where the transmitted symbol is the k-th component of $\vec s_n$\ie $s_{k,n} = [\vec s_n]_k$, 
%the observed noise is $w_{k,n}' = [\ChannelV^\T \vec w_n]_k$,
%$\lambda_{k,n}= [\ChannelD^2]_{kk}$ represents the channel gain and $\sigma_{k,n}^2 = [\PrecoderD^2]_{kk}$ is the transmission power.
%Therefore, in this section we solve the following optimization problem:}

In this context, 
the input-output model $\vec y_{n}' = \ChannelD \PrecoderD \vec s_n + \vec{w}'_n$  can be obtained from the general model in \eqref{eq:Eq_signal_model} by setting $\PrecoderV^{\T} =  \mat I_K $ and $\vec y_{n}'=\ChannelV^\T \vec y_n$. From this, we obtain that the equivalent noise is  $\vec{w}'_n = \ChannelV^\T \vec w_n$. Thus, a set of $K$ independent parallel streams are observed at each channel use. The received signal in the $k$-th stream is 
$ y_{k,n}' = \sqrt{\lambda_{k,n}\sigma_{k,n}^2} s_{k,n} + w_{k,n}',$
where the transmitted symbol is the k-th component of $\vec s_n$\ie $s_{k,n} = [\vec s_n]_k$, 
$w_{k,n}' = [\vec w'_n]_k$ is the observed noise,
$\lambda_{k,n}= [\ChannelD^2]_{kk}$ is the channel gain, and $\sigma_{k,n}^2 = [\PrecoderD^2]_{kk}$ is the transmission radiated power.
Therefore, in this section we solve the following optimization problem:
\begin{equation}
\label{Eq:Problem_Equivalent_Vb_I}
\max_{\sigma_{k,n}^2} \quad \sum_{j = 1}^J \sum_{n \in \Pool} I ( \vec s_n; \vec y_{n}' ), \quad \mathrm{subject~to \quad \eqref{Eq:Problem_Equivalent_ECC}}.
\end{equation}

%where $\bar I (s_{k,n};  y_{k,n}')$ is the input-output mutual information of the scalar channel model in \eqref{Eq:eq_Model}.
Note  that $I(\vec s_n ;  \vec y_{n}') = I(\vec s_n ;  \vec y_n)$ since
a linear unitary rotation in the received signal  does not affect the input-output mutual information \cite{xiao_globallyOptimal_2011}. Thus, the power allocation that maximizes \eqref{Eq:Problem_Equivalent_Vb_I} is equal to the one that maximizes \eqref{Eq:Problem_Equivalent} and it can be obtained by particularizing \eqref{Eq:FirstKKT} with $\PrecoderV^{\T} =  \mat I_K$\ie $\left[ \mat E_n \right]_{kk} = \frac{1}{\lambda_{k,n}W_j}$. From where, it follows that
\begin{sequation}
\label{Eq:Power_alloc_scalar}
\sigma_{k,n}^2 =  \frac{1}{\lambda_{k,n}} mmse_k^{-1}\left( \min\left\{1, \frac{1}{W_j \lambda_{k,n}} \right\} \right) , \quad \forall k, \forall j, \forall n\in\Pool,
\end{sequation}
where $mmse_k^{-1}(\cdot)$ is the inverse \ac{MMSE} function, defined as in \cite{Lozano_mercuryWF_2006}, that returns the \ac{SNR} of the $k$-th stream for a given \ac{MMSE}, which depends on the probability distribution of $s_{k,n}$.

%\section{The \MWFlow[] solution}
%\label{Sec:Interpretation}
%\textcolor{red}{When arbitrary constellations are used, the mutual information asymptotically saturates, e.g., when using BPSK it is not possible to send more than one bit of information per channel use, in opposition to what happens with Gaussian signaling.}
To present the graphical interpretation of the solution, we need to reformulate \eqref{Eq:Power_alloc_scalar} as
%\small
\begin{sequation}
%\textstyle
\label{Eq:Power_alloc_mercury}
\sigma_{k,n}^2 = \left( W_j -  \frac{1}{\lambda_{k,n}} G_k\left(\frac{1}{W_j \lambda_{k,n}}\right)\right)^+ ,\quad \forall k, \forall j, \forall n\in\Pool,
\end{sequation}
%\normalsize
where  $G_k(\psi)$ is defined in \eqref{Eq:G} \cite{Lozano_mercuryWF_2006}, depends on the modulation used, and satisfies the next lemma:
%\vspace{-1em}
\begin{lemma}
\label{L:G_decreasing}
The function $G_k(\psi)$ is monotonically decreasing in $\psi$.
\end{lemma}
%\vspace{-0.8em}
\begin{IEEEproof}
%\vspace{-0.7em}
See Appendix \ref{A:G_decreasing}.
\end{IEEEproof}
\begin{remark}
To demonstrate the validity of the graphical representation presented in this section, we need to analytically demonstrate that $G_k(\psi)$ is monotonically decreasing in $\psi$. 
In \cite{Lozano_mercuryWF_2006}, it was already stated that $G_k(\cdot)$ is decreasing, however, the authors did not provide an analytical proof for their statement. 
Therefore, we consider that Lemma \ref{L:G_decreasing} and its explicit proof are crucial to validate the graphical representation introduced in this section.
\end{remark}

Observe the similarity of the power allocation found in \eqref{Eq:Power_alloc_mercury} with the \MWF in \eqref{Eq:Lozano_mercury} \cite{Lozano_mercuryWF_2006}. The main difference of our solution is that, due to the nature of the energy harvesting process, the water level depends on the channel access. Indeed, from Properties \ref{prop:1} and \ref{prop:emtyBatery} we have seen that the node is able to increase the water level as energy is being harvested.
%In the next two subsections, we give some graphical interpretation of the solution in \eqref{Eq:Power_alloc_mercury}. First, considering arbitrary constellations and, afterwards, particularizing for the capacity achieving Gaussian signals.

Moreover, observe that if we particularize \eqref{Eq:Power_alloc_mercury} for Gaussian distributed inputs, which have  $G_k(\psi) = 1$, $\forall \psi$, (see \cite{Lozano_mercuryWF_2006}), the \ac{DWF} solution in \cite{Ozel_Ulukus_Yener_TX_Fading_Optimal_2011} is recovered.
%Let us particularize \eqref{Eq:Power_alloc_mercury} for the capacity achieving Gaussian distributed inputs. In such a case, $mmse_k(\rho) = (1 + \rho)^{-1}$ and the mercury factor reduces to $G_k(\psi) = 1$, $\forall \psi$ \cite{Lozano_mercuryWF_2006}. Therefore, the optimal power allocation at the $k$-th stream and $n$-th channel use is
%\begin{sequation}
%\label{Eq:Power_alloc_gaussian}
%\sigma_{k,n}^2 = \left( W_j -  \frac{1}{\lambda_{k,n}} \right)^+ .
%\end{sequation}
Therefore, the mercury factor gives a measure of how power allocation is modified when using non-Gaussian input distributions.

%
%\subsection{Mercury Water-Flowing Intuitive Interpretation}

\begin{figure}
\centering
\includegraphics[width=0.8\columnwidth]{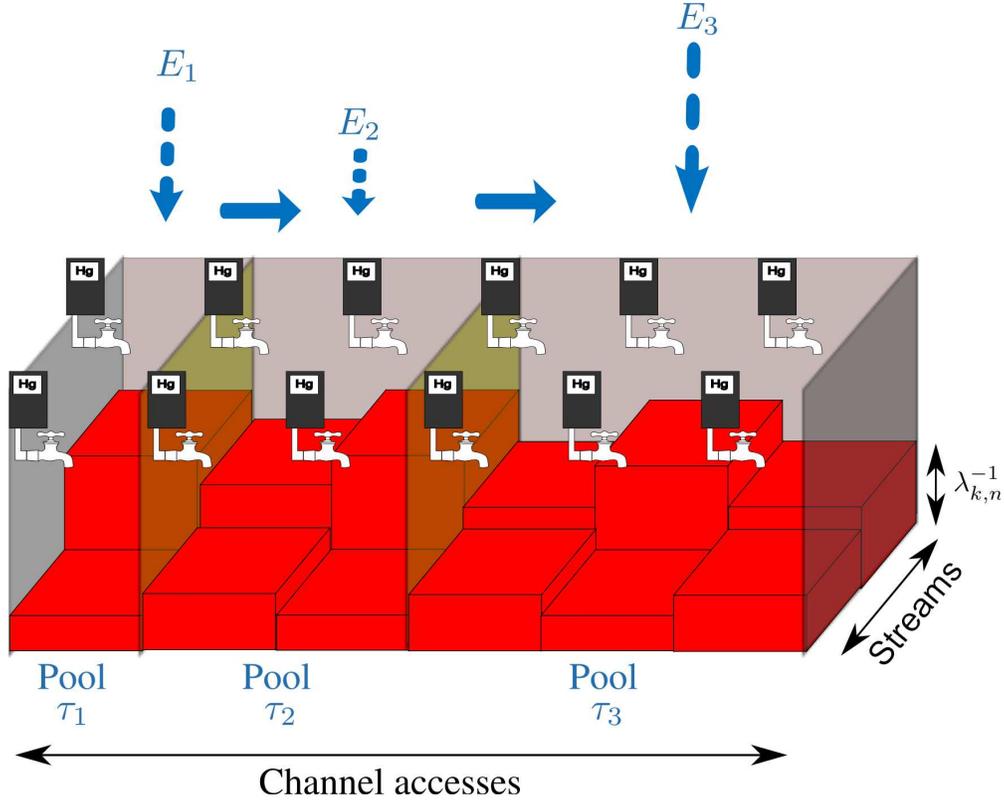}
\caption{Graphical interpretation of the \MWFlow solution, where $N= 6$, $J=3$, and $K=2$.}
\label{fig:interpretation}
\end{figure}

Let $\mathcal{H}_g^{\{k,n\}}(W_j)$ be the mercury level of the $k$-th stream at the $n$-th channel use\ie 
\begin{sequation}
\label{Eq:Mercury_level}
\mathcal{H}_g^{\{k,n\}}(W_j) =  \frac{1}{\lambda_{k,n}}G_k\left(\frac{1}{W_j \lambda_{k,n}} \right), \quad \forall k, \forall j, \forall n\in\Pool,
\end{sequation}
which depends on the gain and water level of the channel use.
Then, the power allocated in a certain stream is the difference between the water and mercury levels\ie $\sigma_{k,n}^2 = \left( W_j - \mathcal{H}_g^{\{k,n\}}(W_j) \right)^+$. The solution interpretation presented in this section is based on the fact that the mercury level is monotonically increasing in $W_j$, which follows directly from Lemma \ref{L:G_decreasing}, and generalizes both the \MWF and the \ac{DWF} solutions derived in \cite{Lozano_mercuryWF_2006} and \cite{Ozel_Ulukus_Yener_TX_Fading_Optimal_2011}, respectively. The \MWFlow interpretation, depicted in \figurename \ref{fig:interpretation}, is the following:

\begin{list}{\labelenumi}{\usecounter{enumi} \leftmargin=0.2cm}
	\item Each parallel channel is represented with a unit-base water-porous mercury-nonporous vessel\footnote{The vessel boundaries are not depicted in \figurename \ref{fig:interpretation} for the sake of simplicity.}.
	\item Then, each vessel is filled with a solid substance up to a height equal to $\lambda_{k,n}^{-1}$.
	\item A water \textit{right-permeable} material is used to separate the different pools.
	\item Each vessel has a faucet that controls the rhythm at which mercury is poured.  The faucet modifies the mercury flow so that the relation between mercury and water levels in \eqref{Eq:Mercury_level} is always satisfied.
	\item Simultaneously,
\begin{list}{\labelitemi}{ \leftmargin=0.3cm}
	\item The water level is progressively increased to all pools at the same time, adding the necessary amount of water to each pool. The maximum amount of water that can be externally added at some pool is given by the pool's harvested power ($E_j/ \Ts$). Let the water freely flow right through the different pools.
	\item Mercury is added to each of the vessels at a different rhythm which is controlled by the vessel's faucet.
\end{list}
	\item The optimal power allocation in each parallel channel is found when all the pools have used all the harvested energy and is obtained as the difference between the water and mercury levels.
\end{list}

%In next section, we introduce two different algorithms that compute the \MWFlow solution.

%With this result, we extend the \textit{directional waterfiling} solution found in \cite{Ozel_Ulukus_Yener_TX_Fading_Optimal_2011} to the \ac{MIMO} scenario. A graphical interpretation can be seen in Figure \ref{fig:interpretation} for the case of having two streams with constant gain along the $N=11$ time slots. Three energy arrivals are produced, where $E_3 > E_1 > E_2$. This energy arrivals are used to fill in their respective pools with water.
%As stated before, the water level is constant along streams and non-decreasing over time. To satisfy this water level property, some water from the pool $\tau_1$ must flow to the pool $\tau_2$ (water can only flow to the right, otherwise, \ac{ECC} would be broken) equalizing the water level in both pools. Moreover observe that no water flow is produce to the pool $\tau_3$ as it already has a higher water level. Once the water level is known, the power allocation is determined by \eqref{Eq:Power_alloc_gaussian}.

%\vspace{-1em}
\section{\MWFlow[] offline algorithms}
\label{Sec:Algorithms}

We have designed two different algorithms to compute the optimal  \MWFlow solution, namely, the \ac{NDA} and the \ac{FSA}, which are presented in Sections \ref{Sec:Algorithm_NDA} and \ref{Sec:Algorithm_FSA}, respectively. Afterwards, in Section \ref{S:Performance}, we prove the algorithms' optimality and analyze their computational complexity.

As shown by the KKT optimality conditions, the water in a certain pool may flow to the pools at its right (i.e., from prior to later time instants). This way, the water level over a consecutive set of  pools may be equalized. This set of constant water level pools is referred to as an \textit{epoch}, $\mathcal E_m, m = 1 \dots M$, where $M$ is the total number of epochs and it is unknown a priori. Note that, since the pools are a partition of the epochs, a certain pool $\tau_j$ is only contained in one epoch. However, an epoch may contain several pools, therefore, $M \leq J$.

To compute the power allocation in \eqref{Eq:Power_alloc_mercury}, we just  need to determine which pools are contained in each epoch $\mathcal E_m$ as, once the epochs are known, the optimal power allocation of the epoch can be found by performing the Mercury/Waterfilling Algorithm (\MWFA) introduced in \cite{Lozano_mercuryWF_2006}, where the $m$-th epoch water level, $\bar W_m$, is found by forcing that the energy expended in the epoch has to be equal to the energy harvested, which follows from Property \ref{prop:emtyBatery}.

The following two algorithms use a different approach two determine the epochs:

%\subsection{\ac{ILA}}
%\label{Sec:Algorithm2}
%This algorithm is based in the interpretation presented in the previous section\ie gradually increase the water and mercury levels in each of the parallel channels. The algorithm proceeds as follows:
%\begin{enumerate}
%	\item Initially, set $M = J$\ie every epoch contains one pool  $\mathcal E_m = \{\tau_m\}$, $m = 1 \dots M$.
%	\item Simultaneously, increase, in a small step, the water level in all the epochs that still have available energy.\label{Alg:increase}  
% 	\item Add mercury to the vessels so that the mercury and water levels satisfy \eqref{Eq:Mercury_level}. 
% 	\item Compute the energy expended in each epoch\footnote{This is done by adding the energy expended in the different epoch's channels, where the energy of a certain channel is obtained by subtracting the channel's water and mercury levels and multiplying by $\Ts$.}. 
%	\item Search for epochs that have used all the epoch available energy. 
%\begin{itemize}
%	\item If no epochs are found, go back to \ref{Alg:increase}. 
%	\item For the epochs found:
%\begin{itemize}
%	\item If the previous epoch still has available energy, merge the two epochs and go back to \ref{Alg:increase}. 
%	\item Otherwise, the epoch has been determined, which means, the optimal power allocation of all the parallel channels contained in the epoch. If the rest of epochs have been also determined, the algorithm ends. Otherwise, go to step \ref{Alg:increase}.
%\end{itemize}
%
%
%\end{itemize}
%	
%\end{enumerate}
%
%

%\vspace{-1.4em}
\subsection{\ac{NDA}}
%\vspace{-0.5em}
\label{Sec:Algorithm_NDA}
The \ac{NDA} uses the fact that a water level decrease is suboptimal, which follows from the KKT conditions (see Property \ref{prop:1}),
to compute the optimal power allocation as follows:

\begin{list}{\labelenumi}{\usecounter{enumi} \leftmargin=0.2cm}
	\item Initially, set $M := J$\ie every epoch contains one pool $\mathcal E_m := \{\tau_m\}$, $m = 1 \dots M.$
	\item Perform the \MWFA in \cite{Lozano_mercuryWF_2006} to every epoch to obtain the water level, $\bar W_m$, in each epoch.
	\item Look for some epoch, $m'$, at which the water level decreases\ie $\bar W_{m'} > \bar W_{m'+1}$: \label{Alg1:Search}
\begin{list}{\labelitemi}{\leftmargin=0.3cm}
    \item If some epoch is found, merge this epoch with the following epoch\ie $\mathcal E_{m'} := \mathcal E_{m'} \cup \mathcal E_{m' + 1 }$. The harvested energy of the resulting epoch is the sum of the two original epochs. Then, the total number of epochs has been reduced by one\ie $M := M - 1$. Perform the \MWFA to obtain the new water level of the $m'$-th epoch\ie $\bar W_{m'}$, and go back to \ref{Alg1:Search}.
	\item  If no epoch is found, the optimal $M$ has been found along with the optimal power allocation.
\end{list}
\end{list}

%\vspace{-1.4em}
\subsection{\ac{FSA}}
%\vspace{-0.5em}
\label{Sec:Algorithm_FSA}
The \ac{FSA} determines the different epochs by finding the optimal \textit{transition pools}, $\{\mathcal T_m ^\star \}_{m=1}^M$, that are defined as the first pool of each epoch. As stated before, once the epochs are known the optimal power allocation is determined by  applying the \MWFA to each epoch.\footnote{Observe that, by definition, $\mathcal T_1^\star$ is the first channel use.} To determine $\{\mathcal T_m ^\star \}_{m=1}^M$, we have designed a forward-search algorithm that extends the algorithm introduced in \cite{ho_optimal_2011} to take into account arbitrary input distributions. We explain how to obtain $\mathcal T_2^\star$ and the others are found in the same manner: 
\begin{list}{\labelenumi}{\usecounter{enumi} \leftmargin=0.2cm}
    \item Assume that the first epoch contains all the pools, $\mathcal E_1 := \{\tau_1,\tau_2 \dots, \tau_J\}$.
    \item Perform the \MWFA in \cite{Lozano_mercuryWF_2006} to the epoch. \label{Step2}
    \item Check whether all the \acp{ECC} within the epoch are fulfilled:
\begin{list}{\labelitemi}{\usecounter{enumii} \leftmargin=0.3cm}
    \item If they are not fulfilled, remove the last pool from $\mathcal E_1$ and go back to step \ref{Step2}.
    \item If they are fulfilled, the optimal transition pool, $\mathcal T_2^\star$, is the first pool not included in the epoch. 
\end{list}
\end{list}

The same procedure is repeated to determine the following transition pools until the $N$-th pool is included in some epoch. When this happens, the optimal power allocation has been found for all the channel accesses and streams.

%\vspace{-1em}
\subsection{Optimality and performance characterization of the offline algorithms}
\label{S:Performance}
In this section, first, we demonstrate the optimality of the \ac{NDA} and the \ac{FSA}, which is presented in Theorem \ref{Th:Optimality_NDA_FSA} and, afterwards, we characterize their associated computational complexity.
%\vspace{-0.4em}
\begin{theorem}
%\vspace{-0.7em}
\label{Th:Optimality_NDA_FSA}
Both the \ac{NDA} and the \ac{FSA} compute the optimal power allocation given in \eqref{Eq:Power_alloc_mercury}.
%\vspace{-5em}
\end{theorem}
\begin{IEEEproof}
See Appendix \ref{Ap:Proof_Theorem}.
\end{IEEEproof}

With the previous theorem, we have demonstrated that both algorithms compute the optimal power allocation, however, the computational cost of such a computation may be very different. To evaluate this, in Appendix \ref{Ap:Complexity}, we have conducted an exhaustive study on the computational complexity of each of these two algorithms. 

%Regarding the \ac{ILA}, we have heuristically observed through simulation that it requires much more computations to determine the solution than the \ac{NDA} and \ac{FSA}, therefore, it has been excluded from the study. 
Our performance analysis is three-fold, namely, the best, worst, and average computational complexities are computed. Note that both algorithms internally call the \MWFA a certain number of times to find the optimal solution. Let  $\CMWF$ denote the number of calls to the \MWFA required to compute the \MWFlow solution, which depends on the algorithm itself and on the dynamics of the energy harvesting process.
In this context, the best or worst computational complexity is the performance when the minimum or maximum number of calls to \MWFA are required, respectively.
The average computational complexity  uses a probabilistic model to compute the average number of calls to \MWFA.
Basically, for the \ac{NDA} we assume that there is  a fixed probability $q$ that the water level decreases from epoch to epoch, whereas,  for the \ac{FSA}  we assume that there is  a fixed probability $p$ that  a certain \ac{ECC} is  not satisfied. Both $p$ and $q$ can be experimentally adjusted depending on the energy harvesting profile. 
The computational complexity in terms of operations (Op.), as well as, in terms of $\CMWF$ is summarized in Table \ref{tab:worst_best}, where $\alpha$ is a constant parameter that depends, among others, on the size of the \ac{MMSE} table required to compute the inverse \ac{MMSE} function and on the tolerance used in the stopping criteria of the $\MWFA$.
The details of the derivations of the different computational complexities can be found in Appendix \ref{Ap:Complexity}. In Section \ref{Sec:Results_algorithms}, the theoretical results on the algorithms' computational complexities are compared with the ones obtained through simulation.

\begin{table*}[t]
%\small
\centering
\begin{tabular}{|c|c |c|| c|c||l|}
\cline{2-6}
\multicolumn{1}{c}{} & \multicolumn{2}{|c||}{Best}  & \multicolumn{2}{|c||}{Worst}  & \multicolumn{1}{|c|}{Average}  \\
\cline{2-6}
\multicolumn{1}{c|}{} & (Op.) & ($\CMWF$) & (Op.) & ($\CMWF$)& \multicolumn{1}{|c|}{($\CMWF$ )}\\
\hline
\multirow{2}{*}{$\ac{NDA}$} & \multirow{2}{*}{$\alpha N K$} &\multirow{2}{*}{$J$}& \multirow{2}{*}{$O(\frac{\alpha}{2} KNJ)$} & \multirow{2}{*}{$2J-1$} & $\mathbb{E}\:\{\CMWF^{NDA}\} = J(q+1)-q$  \\
 & & &  & & $\var \{\CMWF^{NDA}\}= (J-1)q(1-q)$ \\
\hline
\multirow{2}{*}{$\ac{FSA}$} & \multirow{2}{*}{$\alpha N K$} & \multirow{2}{*}{$1$} & \multirow{2}{*}{$O(\frac{\alpha}{6} KNJ^2)$} & \multirow{2}{*}{$\frac{J^2}{2} $} & $\mathbb{E}\:\{\CMWF^{FSA}\} = \left(\frac{J^2}{2} + \frac{J}{2} -1 \right)p +1$ \\ 
 & & & & & $\var \{\CMWF^{FSA}\} = \left(\frac{J}{2}+1\right)^2 (J-1)p(1-p).$ \\
\hline
\end{tabular}
\caption{Computational complexity of the \ac{NDA} and the \ac{FSA} in the best, worst and average case scenarios.}
\label{tab:worst_best}
\end{table*}

%\vspace{-1em}
\section{Online algorithm}
\label{S:online}
Up to now, we have assumed that the transmitter has non-causal knowledge of both the \ac{CSI} and the energy harvesting process, which is not a realistic assumption in practice.
%\ac{CSI} can be provided to the transmitter by means of a feedback link, however, having a complete knowledge of the energy harvesting process is not a realistic assumption in practice. 
Therefore, the \MWFlow solution provides an upper bound on the achievable mutual information of practical schemes in which $\PrecoderV = \mat I_K$. In this section, we develop an online algorithm, which 
is strongly based on the optimal offline solution, the \MWFlow power allocation, but that 
does not require future knowledge of neither the energy arrivals nor  the channel state, that computes a suboptimal power allocation of the problem in  \eqref{Eq:Problem_Equivalent_Vb_I}.

Let $F_w$ be the \textit{flowing window} that is an input parameter of the online algorithm that refers to the number of channel accesses in which the water is allowed to flow, which can be obtained by a previous training under the considered energy harvesting profile, and let an \textit{event} denote  a  channel access in which a change in the channel state is produced or an energy packet is harvested\ie 
$s_t = \{n | \ChannelD[n-1] \neq \ChannelD \} \cup \{ n | n = e_j \}$, $t=1\dots T$, where $T\in [J, N]$.
In this context, the proposed online algorithm proceeds as follows:
(1.) The initial energy in the battery, $E_1$, is allocated to the different streams of the first $F_w$ channel accesses according to the \MWFA where the channel is expected to be static and equal to the gain of the first channel use $\ChannelD = \ChannelD[1], \forall n \in [1, F_w ]$.
%Note that if $e_2 > F_w + 1$, then the transmission power of the slots in $[F_w +1, e_2 -1]$ is zero since by $F_w$ the node has consumed all the available energy; otherwise, if $e_2 \leq F_w +1$, the transmitter  may still have some remaining energy in the battery at $e_2$.
(2.) When the transmitter detects an event, it updates the allocated power of the channel accesses $n \in [s_t, \min \{s_t + F_w -1, N \}]$ by using the \MWFA with the remaining energy in the battery and with the energy of the harvested packet (if the event is an energy arrival)\ie $\sum_{j | s_t \leq  e_j} E_j - \Ts \sum_{n= 1}^{s_t -1} \sum_{k} \sigma_{k,n}^2$, and by assuming that the channel remains constant during the flowing window\ie $\ChannelD = \ChannelD[s_t]$,  $\forall n \in [s_t, \min \{s_t + F_w -1, N \}]$. 
Note that the transmitter may stay silent in some channel accesses if the difference between two consecutive incoming energy packets is greater than the flowing window, $e_j - e_{j-1} > F_w$.\footnote{
This situation rarely takes place in practice since, in most common situations, $F_w$ is several times the mean number of channel accesses per pool. For example, in the simulated framework presented in Section \ref{Sec:Results}, we have obtained that $F_w$ is 4.4 times the mean number of channel accesses per pool.
}
Step (2.) is repeated until the $N$-th channel access is reached. 
The proposed online algorithm satisfies \acp{ECC} and, as pointed out, does not require future information of neither the channel state nor the energy arrivals.

% only requires causal information of the channel state and of the energy harvesting process.
The performance in terms  of achieved mutual information depends on the correctness of the estimation of the flowing window, $F_w$, as discussed with the numerical analysis in Section \ref{Sec:Results}.
In summary, this online algorithm provides us a lower bound on the mutual information that can be achieved with sophisticated online algorithms that make use of precise statistical models of the energy harvesting process and channel state.\footnote{ 
A myriad of works have dealt with channel modeling, however,  having a precise statistical model of the energy harvesting process is indeed not trivial as it depends on many factors such as the harvester used by the node (e.g., a solar panel, piezoelectric generator, etc.), the node's placement, mobility, etc.}

%This online algorithm provides us a lower bound on the mutual information that can be achieved with sophisticated online algorithms that make use of precise statistical models of the energy harvesting process. Note that having a precise statistical model of the energy harvesting process is indeed not trivial as it will depend on many factors such as the harvester used by the node (e.g., a solar panel, piezoelectric generator, etc.), the node's placement, mobility, etc.
\section{Results}
\label{Sec:Results}
This section first evaluates the gain of the proposed \MWFlow solution with respect to other suboptimal solutions and,  secondly, it presents an analysis through simulation of the computational complexity of the \ac{NDA} and \ac{FSA}.

%\vspace{-1em}
\subsection{Results on \MWFlow solution}
\label{Sec:Results_sol}
In this section, we evaluate the  mutual informations obtained with the optimal offline solution, the \MWFlow ($\mathrm{\mathcal H_g}$-$\mathrm{WFlow}$), and with the online policy presented in Section \ref{S:online}.

To the best of our knowledge, there are no  offline algorithms in the literature that maximize the mutual information by jointly considering energy harvesting at the transmitter and arbitrary distributed input symbols.
In this context, we use the following three algorithms,
which are optimal in different setups and have been adapted to the energy harvesting scenario,
 as a reference to evaluate the mutual information achieved  by the  proposed offline and  online  solutions:
(i.) The $\mathrm{DWF}$ solution in \eqref{Eq:Directional_WF} that is the optimal offline power allocation for a \ac{WEHN} when the distribution of the input symbols is Gaussian.
(ii.) Pool-by-Pool Waterfilling ($\mathrm{PbP}$-$\mathrm{WF}$) that uses the \ac{WF} power allocation in \eqref{Eq:Waterfilling} by forcing that the harvested energy in a certain pool is expended in the channel accesses of that same pool.
(iii.) Pool-by-Pool Mercury/Waterfilling ($\mathrm{PbP}$-$\mathrm {\mathcal H_gWF}$)
where the power allocation is obtained by using the \MWF solution in \eqref{Eq:Lozano_mercury} and forcing that the harvested energy in a certain pool is expended in the channel accesses of that same pool.

\begin{figure}[t]
\centering
\includegraphics[width=\columnwidth]{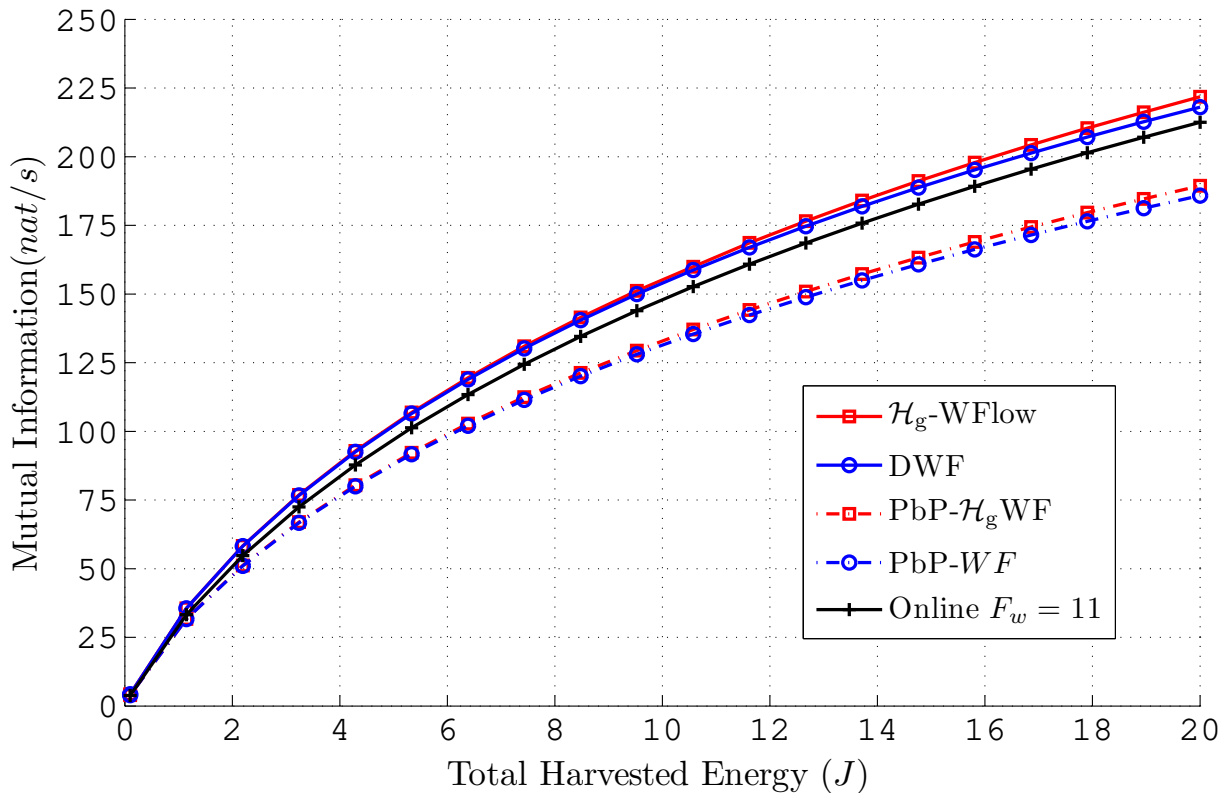}
\caption{Mutual information for the different transmission strategies versus total harvested energy. }
\label{fig:Simu}
%\vspace{-1em}
\end{figure}

We have considered a channel matrix of rank $K=4$, where the channel gains are generated randomly. The modulations used in each stream are BPSK, 4-PAM, 16-PAM, and 32-PAM, respectively. The symbol duration is $\Ts = 10~ms$ and $N=100$ channel accesses have been considered during which a total of $J=40$ energy packets are harvested.
Energy arrivals are uniformly distributed along the channel accesses and with random amounts of energy, which are normalized according to the total harvested energy that varies along the $x$-axis of \figurename \ref{fig:Simu}. The $y$-axis shows the mutual information obtained with the different strategies.
After some training in this scenario, we have obtained that the optimal flowing window is $F_w = 11$ channel accesses.
As shown in \figurename \ref{fig:Simu},  our proposed solution, the $\mathrm{\mathcal H_g}$-$\mathrm{WFlow}$, outperforms all the suboptimal strategies. 
%The improvement of the $\mathrm{\mathcal H_g}$-WFlow and \ac{DWF} \ac{w.r.t.} the \ac{PbP} respective  strategy that are the $\mathrm{PbP}$-$\mathrm {\mathcal H_gWF}$ and the $\mathrm{PbP}$-$\mathrm {WF}$  comes from letting the water to flow across pools and, hence, it directly depends on the parameter $J$ since the higher is the number of pools, the higher is the mutual information gain that can be achieved by letting the water flow.
The improvement of the  $\mathrm{\mathcal H_g}$-$\mathrm{WFlow}$  \ac{w.r.t.} the $\mathrm{PbP}$-$\mathrm {\mathcal H_gWF}$ comes from letting the water to flow across pools and, hence, it directly depends on the parameter $J$ since the higher is the number of pools, the higher is the mutual information gain that can be achieved by letting the water flow.\footnote{
When $J=1$, the solid and dashed curves overlap since there is only one pool.
}
The same happens with the improvement of the $\mathrm{DWF}$ \ac{w.r.t.} $\mathrm{PbP}$-$\mathrm {WF}$.
On the other hand, the mutual information gain of the  $\mathrm{PbP}$-$\mathrm {\mathcal H_gWF}$ and $\mathrm{\mathcal H_g}$-$\mathrm{WFlow}$ \ac{w.r.t.} their respective \ac{WF} strategies, $\mathrm{PbP}$-$\mathrm {WF}$ and $\mathrm{DWF}$, comes from the use of mercury in the resource allocation. 
Thus, when the energy availability is low, both perform similarly because the node is working in the low \ac{SNR} regime in which the mutual information of finite alphabets is well approximated by the mutual information of the Gaussian distribution \cite{Verdu_SE_Wideband_2002}. However, when the energy availability is high, the $\mathrm{PbP}$-\MWF and  $\mathrm{\mathcal H_g}$-$\mathrm{WFlow}$ achieve a higher mutual information than their respective \ac{WF} strategies since
the mutual information of finite constellations asymptotically saturates (not more than $\log_2 Q$ bits of information can be sent per channel use).
Finally, note that, in spite of not having knowledge of the energy arrivals nor channel state, the online power allocation performs close to the 
the offline optimal $\mathrm{\mathcal H_g}$-$\mathrm{WFlow}$ in the low \ac{SNR} regime. 
When the available energy increases, the gap  between the \MWFlow and the proposed online algorithm also increases, nevertheless the online algorithm still presents a reasonably good mutual information outperforming any Pool-by-Pool strategy.

%
%\begin{figure}[t]
%\centering
%\includegraphics[width=1\columnwidth]{Figures/MI_27-Oct-2011_K=4_N=100_J=1.eps}
%\caption{Sum mutual information when no  energy packet is harvested, only the initial battery is used. The pool-by-pool strategies overlap with the  Water-Flowing strategies. The optimal $\mathrm {MIMO \mathcal H_g}$- $\mathrm{WFlowing}$ solution is the red solid line, which, in this setup, is equivalent to the $\mathrm{PbP}$-$\mathrm {\mathcal H_gWF}$.}
%\label{fig:1}
%\end{figure}
The study of the performance in the static scenario is of special interest because the assumption of having future knowledge of the channel state, which has been used for the design of the optimal offline solution, becomes realistic when the channel is static. 
We have evaluated the achieved mutual information in the above setup  for the static channel case and we have obtained similar results than the ones in \figurename \ref{fig:Simu}, where the only difference is that the achieved mutual information   of the different algorithms in the static case is slightly lower since there is less channel gains diversity to assign the available energy.\footnote{
The figure of the static scenario has been omitted for the sake of brevity.}

\begin{figure}[t]
\centering
\includegraphics[width=\columnwidth]{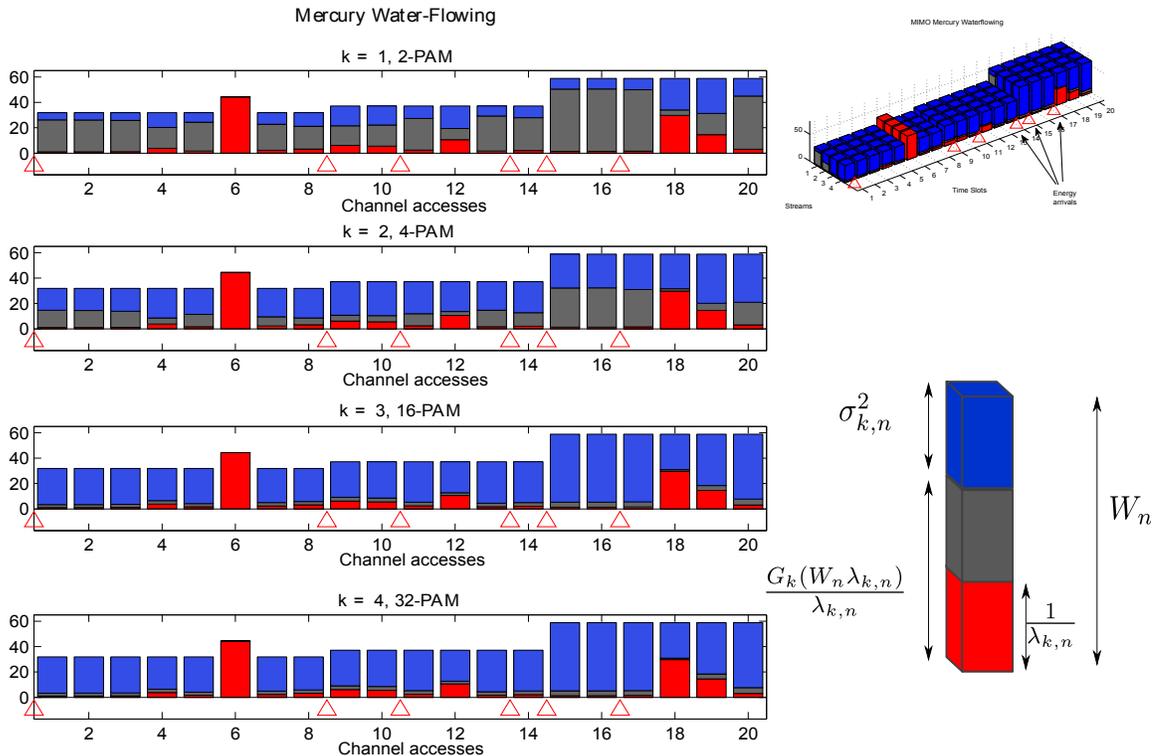}
\caption{Graphical representation of the \MWFlow solution. The red, gray, and blue solid bars represent the inverse of  the channel gain, the mercury and the water levels, respectively. The allocated power is obtained as the difference between the water level and the mercury level.}
\label{fig:wflowing}
%\vspace{-1em}
\end{figure}

In \figurename \ref{fig:wflowing}, the power allocation obtained by the \MWFlow solution in a single simulation is shown for $N=20$ and $K=4$, where the modulations used in the streams 1-4 are BPSK, 4-PAM, 16-PAM and 32-PAM, respectively. Six energy arrivals are produced at the beginning of the channel accesses marked with a triangle. The gains have been generated randomly along channel uses, but fixed constant along streams to ease the observation of the mercury level obtained for the different modulations. As expected from Property \ref{prop:1}, the obtained water level is an increasing step-wise function. Observe that the solution contains three epochs\ie three different water levels, where the pools contained in each epoch are $\mathcal E_1 = \{\tau_1\}$, $\mathcal E_2 = \{\tau_2, \tau_3, \tau_4\}$, and $\mathcal E_3 = \{\tau_5,\tau_6\}$. Moreover, observe that under the same channel gain and water level, the mercury level decreases as the modulation dimension increases.

% This is clearly seen in the different streams of the first channel use, for instance, the mercury level of the first stream, where BPSK is used, is much higher than the one of the fourth stream that uses a 32-PAM modulation. 

%\vspace{-1em}
\subsection{Results on the algorithms' performance}
\label{Sec:Results_algorithms}
In Section \ref{S:Performance}, we have given a summary of the computational complexity of the \ac{NDA} and \ac{FSA} (Table \ref{tab:worst_best} summarizes the obtained results). In this section, we compare the theoretical and experimental performance of both algorithms.

\begin{figure}[t]
\centering
\includegraphics[width=\columnwidth]{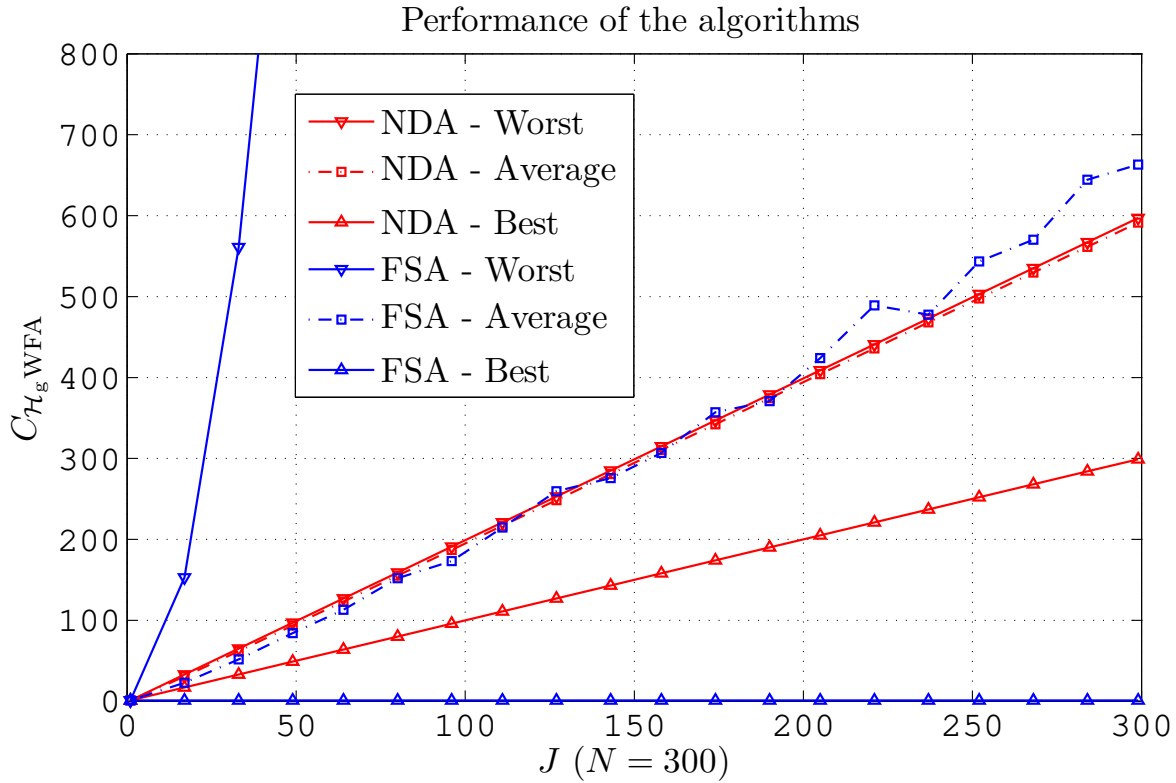}
\caption{Analysis of the performance of the \ac{NDA} and \ac{FSA} in terms of $\CMWF$.}
\label{fig:ALG_performance}
%\vspace{-1em}
\end{figure}

From the simulations, we confirm that, in the best and worst case scenarios, the experimental computational complexity shown in \figurename \ref{fig:ALG_performance} fits the theoretical results presented in Table \ref{tab:worst_best}. Regarding the average case scenario, the mean number of calls to $\MWFA$ of the \ac{NDA} fits the analytical expression $\mathbb{E}\:\{\CMWF^{NDA}\} = J(q+1)-q$ for a value of $q=0.98$. Regarding the  \ac{FSA}, the mean obtained through simulation and the analytically computed expression $\mathbb{E}\:\{\CMWF^{FSA}\} = \left(\frac{J^2}{2} + \frac{J}{2} -1 \right)p +1$ differ from one another. Observe that the quadratic and linear terms of $J$ have the same weight independently of the value of $p$. However, it is easy to observe in \figurename \ref{fig:ALG_performance} that the linear component dominates over the quadratic. Therefore, there is a mismatch between the analytical and experimentally obtained expressions. We believe that this mismatch is due to the fact that in order to obtain some tractable model (see Appendix \ref{Ap:Complexity}), we have assumed that all the \acp{ECC} have the same probability $p$ of not being satisfied, however, in reality this probability is not necessarily equal but depends on the dynamics of the energy harvesting process.

Regardless of the aforementioned mismatch, we observe that, in our simulated energy harvesting set up (the amount of energy in the packets is uniformly distributed), both algorithms have a similar performance in the average case scenario. Note that the difference between the best and worst case scenario is much smaller for the \ac{NDA} than for the \ac{FSA}. This comes from the fact that, in the worst case scenario, the \ac{FSA} has a quadratic dependence in $J$, whereas, for the \ac{NDA} the dependence is linear. This makes the \ac{NDA} more robust in front of changes in the energy harvesting profile.
% Indeed, the \ac{NDA} is working close to its worst performance as $q=0.98$ and still presents the same performance in the average case scenario than the \ac{FSA}.\footnote{Observe that worst case computational complexity is equivalent to the average case computational complexity with $q=1$.} 
In other words, if the energy harvesting profile changes, the \ac{FSA} has more margin to either improve or degrade its performance. For instance, if the node initial battery is very high and the node is operating in the sunset (the amount of harvested energy at the beginning of the transmission duration is higher than the amount harvested at the end) it is likely that the performance of the \ac{FSA} is close to the best case scenario\ie a single call to the \MWFA. On the other hand, if the battery is almost empty at the beginning and the node operates in the sunrise the performance of the \ac{FSA} will be very poor.

To conclude the discussion between the \ac{NDA} and the \ac{FSA}, we want to highlight again that the \ac{NDA} is more robust to changes in the  the energy profile characteristics. However, the \ac{FSA} may be preferable in certain energy harvesting profiles as in its best case performace just requires a call to the \MWFA. Therefore, we believe that the algorithm selection must be done by taking into account the energy harvesting profile and the environmental conditions in which the node is operating.

% he energy packet is then proportia by the sum of the energy that increase proportionally along the x axis that shows the total energy containned in all the packets. 

%Moreover, we want to remark that the throughput differences between the optimal solution, i.e., \textit{MIMO Mercury  Water-Flowing}, and the other two solutions depend both in the number of energy arrivals and the amount of energy contained in the arrivals. For instance, in Figure \ref{fig:0} we show an extreme case where  N=0. Note that if there are no energy arrivals our algorithm does not get any gain with respect to the traditional MIMO waterfilling solution and the curves overlap. Finally, Figure \ref{fig:20} shows the results for $N=20$ and it confirms that the throughput gain of our algorithm increases with the number of energy arrivals. The reason of this gain is that our algorithm saves as much energy as possible by equalizing the power between epochs in comparison with raditional waterfilling. These energy savings are used at the last epochs to speed up the transmission and increase tthe throughput.
%\begin{figure}
%\centering
%\includegraphics[width=0.8\textwidth]{Figures/ch_mimo/0.eps}
%\caption{Throughput comparison when the node does not harvest energy.}
%\label{fig:0}
%\end{figure}
%
%\begin{figure}
%\centering
%\includegraphics[width=0.8\textwidth]{Figures/ch_mimo/20.eps}
%\caption{Throughput comparison for $N=20$ packets of energy.}
%\label{fig:20}
%\end{figure}
%\vspace{-1em}
\section{Conclusions}
\label{Sec:Conclusions}
In this paper, we have considered a \ac{WEHN} transmitting arbitrarily distributed symbols through a discrete-time linear vector Gaussian channel.
We have studied the precoding strategy that maximizes the mutual information by taking into account causality constraints on the use of energy.
We have proved that the optimal left singular vectors of the precoder matrix diagonalize the channel, similarly as in the optimal precoder for the case of non-harvesting nodes.
We have derived the expression $\left[\ChannelD^2 \PrecoderV^\T \MMSE   \PrecoderV \right]_{kk} = \frac{1}{W_j}$ that relates the singular values of the precoder (through the \ac{MMSE} matrix) with the energy harvesting profile (through the different water levels).
The derivation of the optimal right singular vectors, $\PrecoderV^\star$, is left as an open problem.
%Instead, we have derived a framework to numerically obtain the optimal singular values of the precoder given a fixed general $\PrecoderV$.
Then, we have derived the \MWFlow solution, the optimal power allocation when  $\PrecoderV= \mat I_K$, which can be expressed in closed form and accepts an intuitive graphical interpretation based on the fact that the power allocation in a certain stream is the difference between the \WL and the mercury level, which, as shown in this paper,  is a monotonically increasing function of the \WL. 
Additionally, we have developed two different algorithms that compute the \MWFlow solution and we have analytically and experimentally evaluated their computational complexity.
We have also proposed an online algorithm that only requires causal knowledge of the energy harvesting process and channel state.
Finally, through numerical simulations, we have shown a substantial increase in the mutual information \ac{w.r.t.} other suboptimal offline strategies,
which do not account for the shape, size and distribution of the input symbol or  do not exploit the water level equalization across pools, and we have seen that the mutual information achieved with the online algorithm is close to the one of the $\MWFlow$ solution.

\begin{acronym}[OFDMA]

\acro{AWGN}{Additive White Gaussian Noise}
\acro{BCC}{Battery Capacity Constraint}
\acro{BER}{Bit Error Rate}
\acro{CSI}{Channel State Information}
\acro{CWF}{Classical Waterfilling}
\acro{DCC}{Data Causality Constraint}
\acro{DWF}{Directional Water-Filling}
\acro{EBS}{Empty Buffers Strategy}
\acro{ECC}{Energy Causality Constraint}
\acro{FSA}{Forward Search Algorithm}
\acro{ICT}{Information and Communications Technology}
\acro{ICTs}{Information and Communications Technologies}
\acro{IFFT}{Inverse Fast Fourier Transform}
\acro{i.i.d.}{independent and identically distributed}
\acro{ISO}{International Standards Organization}
\acro{ISS}{Incremental Slot Selection}
\acro{LT}{Luby Transform}
\acro{MAC}{Medium Access Control}
\acro{MAP}{Maximum a Posteriori}
\acro{MFSK}{Multiple Frequency-Shift Keying}
\acro{MGSS}{Maximum Gain Slot Selection}
\acro{MI}{Mutual Information}
\acro{MIMO}{Multiple-Input Multiple-Output}
\acro{ML}{Maximum Likelihood}
\acro{MMSE}{Minimum Mean-Square Error}
\acro{NDA}{Non Decreasing water level Algorithm}
\acro{OFDM}{Orthogonal Frequency Division Multiplexing}
\acro{OFDMA}{Orthogonal Frequency Division Multiple Access}
\acro{OSI}{Open System Interconnection}
\acro{PbP}{Pool-by-Pool}
\acro{QoS}{Quality of Service}
\acro{SISO}{Single-Input Single-Output}
\acro{SNR}{Signal to Noise Ratio}
\acro{SVD}{Singular Value Decomposition}
\acro{UPA}{Uniform Power Allocation}
\acro{WEHN}{Wireless Energy Harvesting Node}
\acro{WER}{Word Error Rate}
\acro{WF}{Waterfilling}
\acro{WFlow}{Water-Flowing}
\acro{w.r.t.}{with respect to}
\acro{WSNs}{Wireless Sensor Networks}
\acro{WSS}{Weighted Slot Selection}
\end{acronym}

\appendices
\renewenvironment{seqnarray}
{\noindent \par \vspace{-2em} \small \begin{eqnarray}}
{\end{eqnarray}\normalsize\ignorespacesafterend} 

\setcounter {section} {0}
\def\thesubsection{\Alph{subsection}}
%\vspace{-0.65em}
\section*{Appendix}
\subsection{Proof of Lemma \ref{L:Diag}}
%\vspace{-0.4em}
\label{A:Proof_Diagonalize}
Let us assume that the optimal precoding matrices of the channel accesses $n= 2 \dots N$ are known\ie $\{\Precoder^\star\}_{n=2}^N$. Then, we  focus on finding the optimal precoding matrix of the first channel use $\Precoder[1]^\star$. The problem in \eqref{Eq:P_sloted} is equivalent to
\begin{subequations}
\begin{eqnarray}
%\label{Eq:P_sloted_proof_1}
\max_{\Precoder[1]} && I(\vec s_1; \vec y_1) + a \\
%\end{seqnarray}
%\normalsize
%subject to
%\small
%\begin{seqnarray}
s.t. &&  T_s \Tr(\Precoder[1] \Precoder[1]^\T) + b + c(\ell) \leq \sum_{j=1}^\ell E_j , \quad \ell =1 \dots J, \nonumber
%\label{Eq:P_sloted_proof_3}
\end{eqnarray}
\end{subequations}
where $a$, $b$ and $c(\ell)$ do not depend on $\Precoder[1]$. By only keeping the most restrictive constraint, which is denoted by $P_C$, the previous optimization problem reduces to
\begin{subequations}
\label{Eq:P_proof}
\begin{eqnarray}
\label{Eq:P_proof_1}
\max_{\Precoder[1]} \quad  & & I(\vec s_1; \vec y_1)\\
s.t. \:  \quad
& & \Tr(\Precoder[1] \Precoder[1]^\T) \leq P_C. \label{Eq:P_proof_3}
\end{eqnarray}
\end{subequations}
Finally, once the problem is expressed as \eqref{Eq:P_proof}, it is known from \cite[Prp. 1]{Payaro_Palomar_Optimal_2009} that the left singular vectors of $\Precoder[1]^\star$ can be chosen to coincide with the eigenvectors of $\mat R_{H_1}$\ie $\mat U_{\Precoder[1]} = \mat U_{\mat H_1}$. 
%Hence, the received signal at the first channel use is $\vec y_1 = \mat V_{\mat h_1} \mat \Lambda_1 \mat \Sigma_1 \mat V_{\Precoder[1]}^\T \vec s_1 + \vec w_1$. We define $\vec y_1' = \mat V_{\mat h_1}^\T \vec y_1$ and  $\vec s_1' = \mat V_{\Precoder[1]}^\T \vec s_1$, then, clearly, at the first channel use the channel is diagonalized\ie $\vec y_1' = \mat \Lambda_1 \mat \Sigma_1  \vec s_1' + \vec w_1'$, where $\vec w_1'= \mat V_{\mat h_1}^\T \vec w_1$.
A similar approach can be applied to show that $\{\mat B_n^\star\}_{n=2}^N$ diagonalize their respective channels.
\qed

%\vspace{-1em}
\subsection{Proof of Lemma \ref{L:JacobianI}}
\label{A:JacobianI}
By applying the chain rule, we have that $\D_{\pv}  I(\vec s_n ;   \tilde{\mat H}_n   \vec s_n  +   \vec{w}_n) = \D_{\tilde{\mat H}_n} I(\vec s_n ;   \tilde{\mat H}_n   \vec s_n  +   \vec{w}_n) \:\: \D_{\pv} \tilde{\mat H}_n$. The first term in the previous equation can be easily derived from \eqref{Eq:I_mmse_Palomar} as $\D_{\tilde{\mat H}_n} I(\vec s_n ;   \tilde{\mat H}_n   \vec s_n  + \vec{w}_n) = \v^\T ( \tilde{\mat H}_n \MMSE )$. The second term, $D_{\pv} \tilde{\mat H}_n$, is given in \eqref{Eq:Jacob_pn_Htilde},
\begin{figure*}
\setcounter{equation}{20}
\begin{eqnarray}
D_{\pv} \tilde{\mat H}_n &=&  \frac{1}{2} (\PrecoderV \otimes \ChannelV \ChannelD) \MS \Diag(\vecs \sigma_n^{-1}) 
= \frac{1}{2} (\PrecoderV \otimes \ChannelV \ChannelD) \MS \Diag( \diag (\PrecoderD^{-1}) ) \nonumber \\
&=& \frac{1}{2} (\PrecoderV \otimes \ChannelV \ChannelD) \MS \MS^T (\mat I_K  \otimes \PrecoderD^{-1} ) \MS 
= \frac{1}{2} (\PrecoderV \otimes \ChannelV \ChannelD) (\mat I_K  \otimes \PrecoderD^{-1}) \MS, \label{Eq:Jacob_pn_Htilde}
\end{eqnarray}
% Restore the current equation number.
\setcounter{equation}{21}
% IEEE uses as a separator
%\hrulefill
% The spacer can be tweaked to stop underfull vboxes.
%\vspace{m}
\end{figure*}
where the first equality can be proved in a similar manner than $\D_{\vecs \lambda} \mat P$ in \cite[Proof of Theorem 5]{Payaro_Hessian_2009}. $\MS \in \Re^{K^2 \times K} $ is the reduction matrix introduced in \cite{Payaro_Hessian_2009} (See Appendix \ref{Ap:Prop_reduction} for a concise summary on the properties of $\MS$). In the third and fourth equalities, we have applied Properties 6 and 8 in Appendix \ref{Ap:Prop_reduction}, respectively. Therefore, $\D_{\pv}  I(\vec s_n ;   \tilde{\mat H}_n   \vec s_n  +   \vec{w}_n)$ is derived in \eqref{Eq:DI_1}-\eqref{Eq:DI_4},
\begin{figure*}
% ensure that we have normalsize text
\normalsize
% Store the current equation number.
\vspace{-1em}
\begin{eqnarray}
\D_{\pv}  I(\vec s_n ;   \tilde{\mat H}_n   \vec s_n  +   \vec{w}_n)  &=& \frac{1}{2}\v^\T ( \tilde{\mat H}_n \MMSE) \left(\PrecoderV \otimes \ChannelV \ChannelD\right) \left(\mat I_k \otimes  \PrecoderD^{-1}\right) \MS \label{Eq:DI_1} \\
&=& \frac{1}{2}\v^\T ( \tilde{\mat H}_n \MMSE )  \left(\PrecoderV \otimes \ChannelV \mat \Delta_{n} \PrecoderD^{-1} \right)\MS \label{Eq:DI_2} \\
&=& \frac{1}{2} \v^\T\left( (\ChannelV \mat \Delta_{n} \PrecoderD^{-1})^\T \tilde{\mat H}_n \MMSE \PrecoderV \right) \MS \label{Eq:DI_3} \\
&=& \frac{1}{2} \v^\T \left( \ChannelD^2 \PrecoderV^\T \MMSE   \PrecoderV  \right)\MS =\frac{1}{2} \diag^\T \left( \ChannelD^2 \PrecoderV^\T \MMSE   \PrecoderV  \right),  \label{Eq:DI_4}
\end{eqnarray}
\hrulefill
%\vspace*{4pt}
\end{figure*}
\setcounter{equation}{25}
where, in \eqref{Eq:DI_2} and \eqref{Eq:DI_3}, we have used that 
$(\mat A \otimes \mat B) (\mat C \otimes \mat D) = \mat A \mat C \otimes \mat B \mat D$ and 
$\v( \mat A \mat B \mat C) = (\mat C^\T \otimes \mat A) \v \mat B$
for any matrices $\mat A$, $\mat B$, $\mat C$, and  $\mat D$ such that the matrix products $\mat A \mat C$, $\mat B \mat D$, and  $\mat A \mat B \mat C$ are well defined \cite{Magnus1999}. Finally, \eqref{Eq:DI_4} follows from the definition of the reduction matrix (See Appendix \ref{Ap:Prop_reduction}). This concludes the proof.
\qed

%\vspace{-1em}
\subsection{Proof of Lemma \ref{L:G_decreasing}}
\label{A:G_decreasing}
Let $\psi$ be some fixed \ac{MMSE} that can be obtained as
\begin{sequation}
\label{Eq:mmseGA}
\psi = mmse_G(snr_G) = mmse_A(snr_A),
\end{sequation}
where $mmse_G(snr_G)$ and $mmse_A(snr_A)$ give the \ac{MMSE} as a function of the \ac{SNR} for a Gaussian and for an arbitrary input distribution, respectively. Thus, $snr_G$ and $snr_A$ are the associated required \acp{SNR} to achieve the error $\psi$ for these distributions. 

Similarly, the required \ac{SNR} to obtain a certain error can be computed by the inverse \ac{MMSE} function as
$snr_G = mmse_G^{-1}(\psi)$ and $snr_A = mmse_A^{-1}(\psi)$.

For the Gaussian case, it is broadly known that $\psi = mmse_G(snr_G) =\frac{1}{1 + snr_G}$ with derivative $\frac{\d mmse_G(snr_G)}{\d snr_G} = \frac{-1}{(1 + snr_G)^2}$. Similarly, $snr_G = mmse_G^{-1}(\psi) = \frac{1}{\psi} - 1$ and $\frac{\d  mmse_G^{-1}(\psi)}{\d \psi} = \frac{-1}{\psi^2}$.

Note that for any generic function $f(x)$ it is verified that $\frac{\d f^{-1}(f(x))}{\d x} = \frac{\d f^{-1}(f(x))}{\d f(x)} \frac{\d f(x)}{\d x} =1$. By applying the previous property, the following relation is obtained:
\vspace{-0.6em}
\begin{ssequation}
\frac{\d  mmse_G^{-1}(\psi)}{\d \psi} \frac{\d mmse_G(snr_G)}{\d snr_G} = \frac{\d  mmse_A^{-1}(\psi)}{\d \psi} \frac{\d mmse_A(snr_A)}{\d snr_A}. \nonumber
\end{ssequation}
Recall that  $G(\psi) =\frac{1}{\psi} - mmse_A^{-1}\left( \psi  \right)$ as $\psi \in [0,1]$. Then, its derivative is
\begin{sseqnarray}
\hspace{-0.4cm}\frac{\d G(\psi)}{\d \psi} &=& \frac{-1}{\psi^2} -  \frac{\d  mmse_A^{-1}(\psi)}{\d \psi} \nonumber 			
= \frac{\d  mmse_G^{-1}(\psi)}{\d \psi} -  \frac{\d  mmse_A^{-1}(\psi)}{\d \psi}					 \nonumber \\
&=& \frac{\frac{\d  mmse_G^{-1}(\psi)}{\d \psi}}{\frac{\d mmse_A(snr_A)}{\d snr_A}}  \left( \frac{\d mmse_A(snr_A)}{\d snr_A} - \frac{\d mmse_G(snr_G)}{\d snr_G}\right).  \nonumber
\end{sseqnarray}
In \cite{Guo_estimation_2011}, it was recently shown that  $mmse_A(snr_A) = E\{M_2\}$ and $\frac{\d mmse_A(snr_A)}{\d snr_A}= -E\{M_2^2\}$, where $M_2 = \var \{x| \sqrt{snr_A} x + n\}$. Therefore, the first term of the previous equation is always positive since both the \ac{MMSE} and the inverse \ac{MMSE} functions are decreasing for any distribution. In \eqref{Eq:Sign_1}-\eqref{Eq:Sign_4}, we show that the second term is positive,
\begin{figure*}[t!]
\begin{eqnarray}
\sign \left(\frac{\d G(\psi)}{\d \psi}\right) &=& \sign \left( \frac{\d mmse_A(snr_A)}{\d snr_A} - \frac{\d mmse_G(snr_G)}{\d snr_G}\right)\label{Eq:Sign_1}\\
&=& \sign \left( \frac{\d mmse_A(snr_A)}{\d snr_A} + \frac{1}{(1 + snr_G)^2}\right)										\\
&=& \sign \left( \frac{\d mmse_A(snr_A)}{\d snr_A} + mmse_A(snr_A)^2 \right)				\label{Eq:Sign_3}			\\
&=& \sign \left( -E\{M_2^2\} + \left(E\{M_2\}\right)^2 \right) = \sign \left( - \var \{ M_2 \} \right), \label{Eq:Sign_4}	
\end{eqnarray}
\hrulefill
\end{figure*}
where in \eqref{Eq:Sign_3}, we have used that $snr_G = \frac{1}{mmse_A(snr_A)} -1$, which follows from \eqref{Eq:mmseGA}. In \eqref{Eq:Sign_4}, we have used the recently found expressions of the \ac{MMSE} and its derivative \cite{Guo_estimation_2011}. Finally, as the variance is positive, $\frac{\d G(\psi)}{\d \psi} \leq 0$ and $G(\psi)$ is a monotonically decreasing function. 
\qed

%\vspace{-1em}
\subsection{Proof of Theorem \ref{Th:Optimality_NDA_FSA}}
\label{Ap:Proof_Theorem}
The optimality of the algorithms is proved by demonstrating that the power allocation obtained by means of each of the algorithms satisfies the KKT sufficient optimality conditions:
\renewcommand{\labelenumi}{(\arabic{enumi}.)}
\begin{enumerate}
   \item $\frac{\d \mathcal L}{\d \sigma_{k,n}^2 } = 0$,  $\forall k,n$. \label{KKT:1}
	\item $T_s \sum_{j=1}^\ell \sum_{n\in \Pool} \sum_{k=1}^K \sigma_{k,n}^2 \leq  \sum_{j=1}^\ell E_j $, $\ell=1 \dots J$. \label{KKT:2}
	\item $\rho_\ell \geq 0$, $\ell=1 \dots J$. \label{KKT:4}
	\item $\rho_\ell\big( T_s \sum_{j=1}^\ell \sum_{n\in \Pool} \sum_{k=1}^K\sigma_{k,n}^2  - \sum_{j=1}^\ell E_j \big) = 0$, $\ell=1 \dots J$. \label{KKT:5}
\end{enumerate}
%$	\item $T_s\sum_{n=1}^{N}\sum_{k=1}^K \sigma_{k,n}^2  =  \sum_{j=1}^J E_j H(\ell T_S -e_j)$. \label{KKT:3}$

%\vspace{-2em}
%\small
%\begin{seqnarray}
%\hspace{-0.5cm} T_s\sum_{n=1}^{\ell}\sum_{k=1}^K \sigma_{k,n}^2 \leq  \sum_{j=1}^J E_j H(\ell T_S -e_j) && \quad \forall \ell.\label{Eq:Slackness_proof_1} \\
%\hspace{-0.5cm} T_s\sum_{n=1}^{N}\sum_{k=1}^K \sigma_{k,n}^2  =  \sum_{j=1}^J E_j H(\ell T_S -e_j) && \quad  \label{Eq:Slackness_proof_2}\\
%\hspace{-0.5cm}\rho_\ell \geq 0, &&\quad \forall \ell\label{Eq:Slackness_proof_3},\\
%\hspace{-0.5cm}\rho_\ell\big( T_s\sum_{n=1}^{\ell}\sum_{k=1}^K\sigma_{k,n}^2  - \sum_{j=1}^J E_j H(\ell T_S -e_j)\big) = 0, && \quad \forall \ell.\label{Eq:Slackness_proof_4} \\
%\frac{\d \mathcal L}{\d \sigma_{k,n}^2 } = 0 && \quad \forall k,n.\label{Eq:Slackness_proof_5} 
%\end{seqnarray}
%\normalsize
Moreover, we know that by the end of the transmission the battery must be empty since, otherwise, the remaining energy in the battery can be used to increase the total mutual information.
Thus, (2.) must be met with equality for $\ell = J$.
Note that both algorithms compute a power allocation strategy that satisfies \acp{ECC} and that by the end of the last channel access all the energy has been used. Therefore, (2.) is satisfied $\forall \ell$ and it is satisfied with equality for $\ell =J$. From Property \ref{prop:1}, if the water level is non-decreasing in time then (3.) can be verified. In the \ac{NDA}, the water level is clearly non-decreasing in time. Regarding the \ac{FSA}, if some \ac{ECC} is not satisfied, it is because the water level must be reduced before the point where the \ac{ECC} is not satisfied and increased afterwards. Indeed, this is what the algorithm does in the procedure of finding the optimal epochs. Therefore, (3.) is also satisfied in the \ac{FSA}. Finally, since both algorithms compute the optimal power allocation within an epoch by using the $\MWFA$, where the water level is found by forcing that all the available energy must be used by the end of the epoch,  conditions (1.) and (4.) are satisfied. With this, we have demonstrated that the power allocation computed by the \ac{NDA} and the \ac{FSA} is the optimal power allocation. \qed

\subsection{Computational complexity of the algorithms}
\label{Ap:Complexity}
In this appendix, we study the performance of the two algorithms that compute the \MWFlow solution, the \ac{NDA} and the \ac{FSA}.

We have carried out a three-fold analysis, namely, the best, worst and average computational complexity.
As mentioned before, both algorithms internally call the \MWFA a certain number of times to find the optimal solution. 
The performance is evaluated in terms of operations and number of calls to the \MWFA required to compute the \MWFlow solution, $\CMWF$.

Before getting into the complexity of each of the aforementioned scenarios, let us first compute the complexity of the $\mathrm{\mathcal H_gWFA}$ when the algorithm computes the power allocation of $NK$ parallel channels, where $N$ and $K$ denote the number of channel accesses and  streams, respectively\ie
\begin{equation}
\label{Eq:Complexity_HgWF}
\CCMWF(N, K) = \alpha N K,
\end{equation}
where $\alpha$ is a constant parameter that depends, among others, on the size of the \ac{MMSE} table required to compute the inverse mmse function $mmse_k^{-1}(\cdot)$ and on the tolerance used in the stopping criteria of the $\MWF$.
 Now, let us proceed to compute the computational complexity of the \ac{NDA} and \ac{FSA}.

%First, we have evaluated the computational complexity in the worst case scenario\footnote{By worst case scenario, we mean that to determine the solution each algorithm will call the $\mathrm{\mathcal H_gWFA}$ the maximum number of times it can be called, which depends on the algorithm itself.} for each of the algorithms. Second, a probabilistic analysis of the number calls to Mercury/Waterfilling algorithm has been done for both algorithms, deriving the mean and variance of the number of calls to Mercury/Waterfilling algorithm. 
\subsubsection{Computational complexity in the best case scenario}
$ $ 

\textbf{\ac{NDA}:}
The best case scenario for the \ac{NDA} occurs when the resulting \WLs of applying  the $\mathrm{\mathcal H_gWFA}$ at each pool are non-decreasing throughout all the transmission. Thus, the best case computational complexity for the \ac{NDA} is 
\begin{equation}
%\label{}
CC^B_{NDA}(N,K,J) = \sum_{j=1}^{J} \CCMWF(L_j, K) = \sum_{j=1}^{J}\alpha L_j K = \alpha N K,
\end{equation}
where $L_j$ is the number of channel accesses contained in $\Pool$ and, accordingly, $\sum_{j=1}^{J} L_j =N$. Note that the number of calls to the $\MWF$ is $\CMWF=J.$

\textbf{\ac{FSA}:}
Regarding the \ac{FSA} the best performance is obtained when the algorithm can stop at the first iteration\ie after applying the  $\mathrm{\mathcal H_gWFA}$ to the $N$  channel accesses it is observed that the resulting power allocation satisfies  all energy causality constraints\ie
\begin{equation}
%\label{}
CC^B_{FSA}(N,K,J) = \CCMWF(N, K) = \alpha N K.
\end{equation}
Note that the number of calls to the $\MWF$ for the \ac{FSA} in the best case scenario is $\CMWF=1$.

Observe that, even though $\CMWF$ differs from one algorithm to another one, they achieve the same computational complexity in terms of operations in the best case scenario. However, note that the best case scenario for the \ac{FSA} occurs when the water level of the optimal power allocation remains constant throughout all the transmission time, in other words, there is a single \textit{epoch}. However, the best case scenario for the \ac{NDA} is completely the opposite, the water level is different at every pool and, thus, the total number of epochs is $J$.

\subsubsection{Computational complexity in the worst case scenario}
$ $

\begin{figure}
\centering
\includegraphics[width=0.5\textwidth]{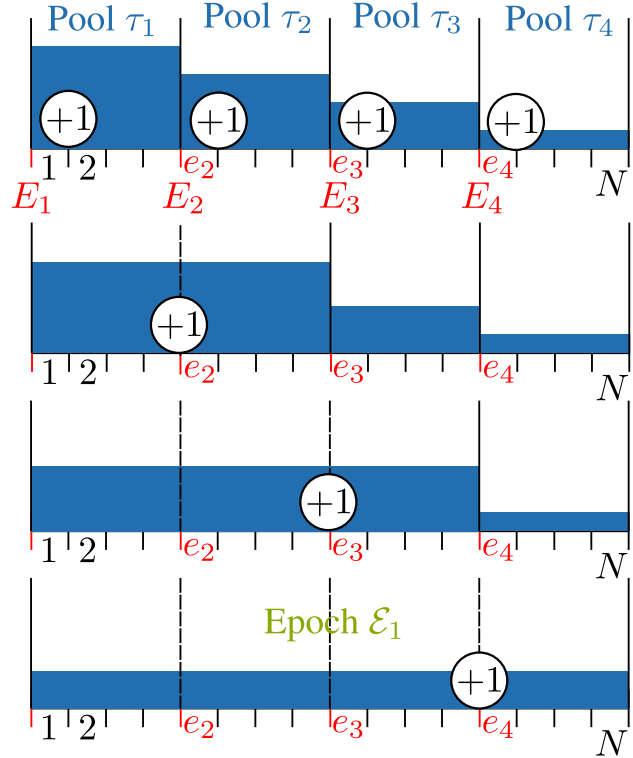}
\caption{Representation of the \ac{NDA} algorithm.}
\label{fig:NDA}
\end{figure}

\textbf{\ac{NDA}:}
The worst case computational complexity for the \ac{NDA} is produced when at every iteration of the algorithm it is observed that the water level is decreasing in some pool transition. \figurename \ref{fig:NDA} shows an example of how the algorithm proceeds for $J=4$. In the first iteration a total of $J$ calls to $\MWF$ are required. Then, in the second iteration, an additional call is performed to merge the first two epochs where it is observed that the \WL is decreasing. As we are considering the worst case scenario, the resulting \WLs will be decreasing at some epoch transition and an additional call is required until all pools have been merged in a single epoch, therefore, the worst case computational complexity for the \ac{NDA} is 
\begin{eqnarray}
CC^W_{NDA}(N,K,J) &=& \sum_{j=1}^{J} \CCMWF(L_j, K) +  \sum_{j=2}^{J} \CCMWF(jL_j, K) \\
&=& \alpha KN + \sum_{j=2}^{J} \alpha K j N/J  \label{Eq:worst_NDA}\\
&=& O(\frac{\alpha}{2} KNJ),
\end{eqnarray}
where the first summation comes from the first iteration of the algorithm and the second one comes from merging the pools with decreasing water level\ie iterations from $2$ to $J$. In \eqref{Eq:worst_NDA}, we have made the simplification of having equal length pools\ie $L_j =N/J$, $\forall j$. The number of calls to $\MWF$ is $\CMWF = 2J-1$.

\textbf{\ac{FSA}:}
The \ac{FSA} starts by assuming that the first epoch contains all the pools, then, it performs $\MWF$ and checks whether the energy causality constraints are satisfied, which are not as we are considering the worst case scenario. Then, it removes the last pool from $\Epoch[1]$ and tries again and so forth until $\Epoch[1]$ just contains one pool and then the constraints must be satisfied. Therefore, a total of $J$ iterations are required to determine $\Epoch[1]^\star$. Similarly, $J-1$ iterations are required to determine $\Epoch[2]^\star$. The computational complexity at each iteration is summarized in Table \ref{tab:worst_FSA} from where we can conclude that the worst case computational complexity of the \ac{FSA} is
\begin{eqnarray}
CC^W_{FSA}(N,K,J) &=& \sum_{j=1}^{J} \alpha K L j (J - j + 1)  \\
&=& \alpha K  \frac{N}{J} \sum_{j=1}^J j J -j^2 +j = O(\frac{\alpha}{6} KNJ^2) \label{Eq:worst_FSA},
\end{eqnarray}
where in \eqref{Eq:worst_FSA} we have made the simplification of having equal length pools\ie $L_j =N/J$, $\forall j$. As every iteration performs a call to $\MWF$, the total number of calls is $\CMWF = \frac{J(J+1)}{2}$.

\begin{table}[t]
\centering
\begin{tabular}{c|c| c|}
\hline
\multicolumn{1}{|c|}{Iteration} & Epoch & Complexity \\
\hline
\multicolumn{1}{|c|}{$1$} & $\Epoch[1] =\{\Pool[1], \dots, \Pool[J]\}$ & $\alpha K \sum_{j=1}^J L_j$ \\
\multicolumn{1}{|c|}{$2$} & $\Epoch[1] =\{\Pool[1], \dots, \Pool[J-1]\}$ & $\alpha K \sum_{j=1}^{J-1} L_j$ \\
\multicolumn{1}{|c|}{$\dots$} & $\dots $ & $\dots$ \\
\multicolumn{1}{|c|}{$J$} & $\Epoch[1] =\{\Pool[1]\}$ & $\alpha K L_1$ \\
%\hline
\hline
\multicolumn{2}{c|}{} & Total $\Epoch[1]^\star = \alpha K [J L_1 + (J-1) L_2 + \dots + L_J]$ \\
\cline{3-3}\noalign{\vspace{0.3cm}}
\hline
\multicolumn{1}{|c|}{$J+1$} & $\Epoch[2] =\{\Pool[2], \dots, \Pool[J]\}$ & $\alpha K \sum_{j=2}^J L_j$ \\
\multicolumn{1}{|c|}{$\dots$} & $\dots $ & $\dots$ \\
\multicolumn{1}{|c|}{$2J-1$} & $\Epoch[2] =\{\Pool[2]\}$ & $\alpha K L_2$ \\
\hline
\multicolumn{2}{c|}{}& Total $\Epoch[2]^\star = \alpha K [(J-1) L_2 +  (J-2) L_3 + \dots + L_J]$ \\
\cline{3-3}
\multicolumn{3}{c}{ $\dots$}\\
\hline
\multicolumn{1}{|c|}{$\frac{J(J+1)}{2}$} & $\Epoch[J] =\{\Pool[J]\}$ & $\alpha K  L_J$ \\
\hline
\multicolumn{2}{c|}{}& Total $\Epoch[J]^\star = \alpha K  L_J$ \\
\cline{3-3}
\end{tabular}
\caption{Computational complexity of the \ac{FSA} in the worst case scenario.}
\label{tab:worst_FSA}
\end{table}

\subsubsection{Computational complexity in the average case scenario}
For the average case scenario, due to the inherent difficulty of determining the computational complexity measured in operations, we have just derived  the complexity in terms of calls to the $\mathrm{\mathcal H_gWFA}$\ie $\CMWF$. By doing this, we can see how the computational complexity is affected by the number of energy arrivals $J$.

\textbf{\ac{NDA}:}
We start by analyzing the average performance of the \ac{NDA}.
Let $q_j$, $j=1\dots J-1$, be the probability that the water-level decreases at some pool transition. Let us assume equal probability at all pool transition $q_j=q$, $\forall j$. Let $\CMWF^{NDA}$ be a random variable that, for a certain call to the \ac{NDA} algorithm, denotes the number of calls to the $\mathrm{\mathcal H_gWFA}$. Note that the minimum number of calls to the $\mathrm{\mathcal H_gWFA}$ is  J and, from here, an additional call is produced every time that a water level decrease is produced. Observe that this additional number of calls is a binomial distribution of parameters $J-1$ and $q$\ie $\mathbb {B}\:(J-1,q)$. Therefore, $\CMWF^{NDA} =J + \mathbb B \:(J-1,q)$ and the mean and variance are
\begin{eqnarray}
%\label{}
\mathbb{E}\:\{\CMWF^{NDA}\} &=& J + \mathbb{E}\{\mathbb {B}\:(J-1,q)\} = J+(J-1)q = J(q+1)-q, \\
\var \{\CMWF^{NDA}\} &=&  \var\{ \mathbb {B} \:(J-1,q)\}= (J-1)q(1-q).
\end{eqnarray}

\textbf{\ac{FSA}:}
Similarly for the \ac{FSA}, let $p_j$,  $j=1\dots J-1$, denote the probability that the $j$-th energy causality constraint of the $FSA$ is not satisfied. We assume that this probability is equal for all the constraints $p_j=p$, $\forall j$. Let $\CMWF^{FSA}$ be a random variable that, for a certain call to the \ac{FSA} algorithm, denotes the number of calls to the $\MWFA$. To determine $\mathbb{E}\:\{\CMWF^{FSA}\}$ for a general $J$, let us first obtain $\CMWF^{FSA}$ for some specific values of $J$ as a function of the broken constraints. Note that up to $J-1$ constraints can be broken. In Tables \ref{tab:worst_FSA_3}, \ref{tab:worst_FSA_4}, \ref{tab:worst_FSA_5}, \tickYes and \tickNo denote that a certain constraint is satisfied or broken, respectively. For example, Table \ref{tab:worst_FSA_3} shows $\CMWF^{FSA}$ when $J=3$ and the energy constraints that can be broken are in the transitions of $\Pool[1] \to \Pool[2]$, which is depicted in the first column, and $\Pool[2] \to \Pool[3]$, in the second column. Similarly, in Tables \ref{tab:worst_FSA_4} and \ref{tab:worst_FSA_5} show the obtained values of $\CMWF^{FSA}$ for $J=4$ and $J=5$, respectively. After carefully examining the previous tables, one may realize that there exists a fixed cost that depends on the number of broken constraints $b$ that is $b + 1$ (at least, one call to $\MWF$ is required before and after the broken constraint) and a variable cost that depends on the placement of the broken constraint. If the broken constraint is the last one the variable cost is $1$. If it  is the one before the last one, the variable cost is $2$ and so forth up to the case in which the broken constraint is the first energy causality constraint where the variable cost is $J-1$. From this observation we can find $\mathbb{E}\:\{\CMWF^{FSA}\}$ for a general $J$ as 
\begin{eqnarray}
\mathbb{E}\:\{\CMWF^{FSA}\} &=&\sum_{b=0}^{J-1} \left[ \binom{J-1}{b} (b+ 1) + \binom{J-2}{b-1} \frac{(J-1)J}{2} \right] p^b (1-p)^{J-1-b}\\
 &=& \sum_{b=0}^{J-1} \binom{J-1}{b}\left( b\left(\frac{J}{2}+1\right) + 1\right)p^b (1-p)^{J-1-b}, \\
 &=& \left(\frac{J}{2}+1\right)(J-1)p +1 = \left(\frac{J^2}{2} + \frac{J}{2} -1 \right)p +1, \label{Eq:FSA_A}
\end{eqnarray}
where in \eqref{Eq:FSA_A}, we have used that the mean of a binomial distribution with parameters $n$ and $p$ is $np$. Similarly, the variance of $\CMWF^{FSA}$ can be obtained through the variance of  a binomial distribution as 
\begin{equation}
%\label{}
\var \{\CMWF^{FSA}\} = \left(\frac{J}{2}+1\right)^2 (J-1)p(1-p).
\end{equation}

This concludes the analysis of the computational complexity of the algorithms.

\begin{table}[t]
\centering
\begin{tabular}{|c c|c|c|}
\hline
\multicolumn{4}{|c|}{J=3} \\
\hline
\hline
\multicolumn{2}{|c|}{Constraint} & $\CMWF^{FSA}$ & Probability\\
\hline
\tickYes & \tickYes  & 1 & $(1 - p)^2$ \\
\tickYes & \tickNo  & 3 & $(1 - p)p$ \\
\tickNo & \tickYes  & 4 & $(1 - p)p$\\
\tickNo & \tickNo  & 6 & $p^2$\\
\hline
\hline
\multicolumn{4}{|c|}{$\mathbb{E}\:\{\CMWF^{FSA}\} = (1 - p)^2 + 7 (1 - p)p +6p^2 $}\\
\hline
\end{tabular}
\caption{Computational complexity of the \ac{FSA} in the average case scenario (in terms of calls to $\MWF$) for $J=3$.}
\label{tab:worst_FSA_3}
\end{table}
\begin{table}[t]
\centering
\begin{tabular}{|c c c|c|c|}
\hline
\multicolumn{5}{|c|}{J=4} \\
\hline
\hline
\multicolumn{3}{|c|}{Constraint} & $\CMWF^{FSA}$ & Probability \\
\hline
\tickYes & \tickYes & \tickYes & 1 & $(1 - p)^3$ \\
\tickYes & \tickYes & \tickNo  & 3 & $(1 - p)^2p$ \\
\tickYes & \tickNo & \tickYes  & 4 & $(1 - p)^2p$\\
\tickNo & \tickYes &  \tickYes  & 5 & $(1 - p)^2p$\\
\tickYes & \tickNo & \tickNo  & 6 & $(1-p)p^2$\\
\tickNo & \tickYes & \tickNo  & 7 & $(1-p)p^2$\\
\tickNo & \tickNo  & \tickYes & 8 & $(1-p)p^2$\\
\tickNo & \tickNo  & \tickNo & 10 & $p^3$\\
\hline
\hline
\multicolumn{5}{|c|}{$\mathbb{E}\:\{\CMWF^{FSA}\} = (1 - p)^3 + 12 (1 - p)^2p + 21 (1-p)p^2 + 10p^3 $}\\
\hline
\end{tabular}
\caption{Computational complexity of the \ac{FSA} in the average case scenario (in terms of calls to $\MWF$) for $J=4$.}
\label{tab:worst_FSA_4}
\end{table}

\begin{table}[t]
\centering
\begin{tabular}{|c c c c|c|c|}
\hline
\multicolumn{6}{|c|}{J=5} \\
\hline
\hline
\multicolumn{4}{|c|}{Constraint} & $\CMWF^{FSA}$ & Probability \\
\hline
\tickYes & \tickYes & \tickYes & \tickYes & 1 & $(1 - p)^4$ \\
\tickYes &\tickYes & \tickYes & \tickNo  & 3 & $(1 - p)^3p$ \\
\tickYes & \tickYes & \tickNo & \tickYes  & 4 & $(1 - p)^3p$\\
\tickYes &\tickNo & \tickYes &  \tickYes  & 5 & $(1 - p)^3p$\\
\tickNo &\tickYes & \tickYes &  \tickYes  & 6 & $(1 - p)^3p$\\
\tickYes &\tickYes & \tickNo & \tickNo  & 6 & $(1-p)^2p^2$\\
\tickYes &\tickNo & \tickYes & \tickNo  & 7 & $(1-p)p^2$\\
\tickNo &\tickYes & \tickYes & \tickNo  & 8 & $(1-p)^2p^2$\\
\tickYes &\tickNo & \tickNo  & \tickYes & 8 & $(1-p)^2p^2$\\
\tickNo &\tickYes & \tickNo  & \tickYes & 9 & $(1-p)^2p^2$\\
\tickNo &\tickNo & \tickYes  & \tickYes & 10 & $(1-p)^2p^2$\\
\tickYes & \tickNo & \tickNo  & \tickNo & 10 & $(1-p)p^3$\\
\tickNo & \tickYes & \tickNo  & \tickNo & 11 & $(1-p)p^3$\\
\tickNo & \tickNo & \tickYes  & \tickNo & 12 & $(1-p)p^3$\\
\tickNo & \tickNo & \tickNo  & \tickYes & 13 & $(1-p)p^3$\\
\tickNo & \tickNo & \tickNo  & \tickNo & 15 & $p^4$\\
\hline
\hline
\multicolumn{6}{|c|}{$\mathbb{E}\:\{\CMWF^{FSA}\} = (1 - p)^4 + 18 (1 - p)^3p + 48 (1-p)^2p^2 + 46 (1-p)p^3 + 15p^4 $}\\
\hline
\end{tabular}
\caption{Computational complexity of the \ac{FSA} in the average case scenario (in terms of calls to $\MWF$) for $J=5$.}
\label{tab:worst_FSA_5}
\end{table}

\subsection{Properties of the reduction matrix}
\label{Ap:Prop_reduction}
The reduction matrix, $\MS \in \Re^{K^2\times K}$, was introduced in \cite{Payaro_Hessian_2009} and is defined as:
\begin{equation}
\label{}
[\MS]_{i+(j-1)k,z} =\delta_{ijz}, \quad \{i,j,z\}\in [1,k]
\end{equation}
Note that from the structure of $\MS$, in each column there is only one entry different than zero and it is equal to one. For instance, the matrices for $K=2$ and $K=3$ are:
\[
 \MS[2] =
 \begin{pmatrix}
  1 & 0  \\
  0 & 0  \\
  0 & 0  \\
  0 & 1  \\
 \end{pmatrix},
\quad
\mathrm{and} \quad
 \MS[3] =
 \begin{pmatrix}
  1 & 0 & 0\\
  0 & 0 & 0 \\
  0 & 0 & 0  \\
  0 & 0 & 0  \\
  0 & 1 & 0 \\
  0 & 0 & 0  \\
  0 & 0 & 0  \\
  0 & 0 & 0  \\
  0 & 0 & 1  \\
 \end{pmatrix}
.\]
The reduction matrix is designed so that 
\begin{equation}
%\label{}
\MS^\T \v(\mat A) = \diag (A)
\end{equation}
for $\mat A\in \Re^{K\times K}$.
In this appendix, we summarize some  additional properties of the reduction matrix:

\begin{prop}
\label{Prop:Multiplication}
Multiplication properties:
\begin{itemize}
    \item Let $\mat A \in \Re^{K^2 \times R}$, then the multiplication $\MS^\T \mat A$ removes $K^2 -K$ rows of $\mat A$.
    \item Let $\mat A \in \Re^{K \times R}$, then the multiplication $\MS \mat A$  adds $K^2 -K$ rows of zeros to $\mat A$.
    \item Let $\mat A \in \Re^{R\times K}$, then the multiplication $\mat A \MS^\T$  adds $K^2 -K$ columns of zeros to $\mat A$.
    \item Let $\mat A \in \Re^{R\times K^2}$, then the multiplication $\mat A \MS$  removes $K^2 -K$ columns of $\mat A$.
\end{itemize}
\end{prop}
\begin{IEEEproof}
The proof follows directly from the structure of the reduction matrix.
\end{IEEEproof}

\begin{prop}
\label{P:MS_Kron_Had}
 Let $\mat A \in \Re^{K \times R}$, $\mat B \in \Re^{K \times R}$, then $\MS^\T \left( \mat A \otimes \mat B \right) \MS = \mat A \circ \mat B$.
\end{prop}
\begin{IEEEproof}
See \cite[Lemma A.2]{Payaro_Hessian_2009}.
\end{IEEEproof}

\begin{prop}
$\MS^\T \MS =\mat I_K$.
\end{prop}
\begin{IEEEproof}
The proof directly follows from setting $ \mat A = \mat I_K$ and $\mat B = \mat I_K$ in Property \ref{P:MS_Kron_Had}.
\end{IEEEproof}

\begin{prop}
Let $\mat A \in \Re^{K \times K}$, then $\MS^\T (\mat A \otimes \mat I_K) \MS = \Diag(\diag(A))$.
\end{prop}
\begin{IEEEproof}
The proof directly follows from setting $\mat B = \mat I_K$ in Property \ref{P:MS_Kron_Had}.
\end{IEEEproof}

\begin{prop}
Let $\vec v \in \Re^{K}$, then $\MS^\T (\vec v \otimes \mat I_K) = \Diag(\vec v)$.
\end{prop}
\begin{IEEEproof}
The Kronecker product expands the vector $\vec v$ in a $K^2 \times K$ matrix that stacks $K$ diagonal matrices. Then, the multiplication by  $\MS^\T$ eliminates rows (see Property \ref{Prop:Multiplication}) so that the resulting matrix is $\Diag(\vec v)$.
\end{IEEEproof}

\begin{prop}
Let $\mat A \in \Re^{K^2 \times K^2}$ be a diagonal matrix, then $\MS \MS^\T \mat A \MS = \mat A \MS$
\end{prop}
\begin{IEEEproof}
From Property \ref{Prop:Multiplication}, $\MS^\T \mat A$ removes rows from $\mat A$. Then, the product by the left by $\MS$ adds rows of zeros. As a result, $\MS \MS^\T \mat A \in \Re^{K^2 \times K^2}$ zeroes $K^2 - K$ rows of $\mat A$. Finally, the product with \MS from the right removes $K^2 - K$ columns. As $\mat A$ is diagonal, the entries that are modified by multiplying from the left by  $\MS \MS^\T$ are later removed by multiplying from the right by $\MS$. Therefore, $\MS \MS^\T \mat A \MS$ is equal than $\mat A \MS$, which directly removes the columns.
\end{IEEEproof}

\bibliographystyle{IEEEtran}

\bibliography{Bib_Maria}

\end{document}